%% file: main_arxiv.tex
\documentclass[reprint, superscriptaddress, aps,prl,twocolumn]{revtex4-1} 
\usepackage{graphicx}
\usepackage{amsmath,amssymb}
\usepackage{mathtools}
\usepackage{multirow}
\usepackage{textcomp}
\usepackage{float}
\usepackage{color}
\usepackage{xcolor}
\usepackage{pdfpages}
\usepackage{ulem}
\usepackage{dsfont}
\usepackage{upgreek}
\usepackage{comment}
\usepackage{hyperref}

\hypersetup{colorlinks, linkcolor={blue!80!black}, citecolor={blue!80!black}, urlcolor={blue!80!black}}

\makeatletter
\AtBeginDocument{\let\LS@rot\@undefined}
\makeatother

\definecolor{bluegreen}{HTML}{16b5b2}

\newcommand{\avg}[1]{\langle #1 \rangle}
\newcommand{\Htwist}{\ensuremath{H_\textrm{twist}}}
\newcommand{\Hct}{\ensuremath{H_\textrm{ct}}}

\newcommand{\uvec}[1]{\ensuremath{\hat{\mathbf{#1}}}}
\newcommand{\abs}[1]{\ensuremath{\left\vert{#1}\right\vert}}

\renewcommand{\vec}[1]{\ensuremath{\mathbf{#1}}}

\newcommand{\Hctg}{\ensuremath{H_{\mathrm{gct}}}}
\renewcommand{\k}{\ensuremath{\vec{k}}}
\newcommand{\0}{\ensuremath{\vec{0}}}
\newcommand{\tJ}{\ensuremath{\tilde{J}}}
\newcommand{\kmin}{\ensuremath{k_c}}
\newcommand{\ginv}{\ensuremath{\chi}}
\newcommand{\Smax}{\ensuremath{S^\mathrm{max}}}
\newcommand{\Smin}{\ensuremath{S^\mathrm{min}}}
\newcommand{\Cmax}{\ensuremath{C^\mathrm{max}}}
\newcommand{\Jtot}{\ensuremath{J_\mathrm{tot}}}
\newcommand{\phisq}{\ensuremath{\phi_{\mathrm{sq}}}}
\newcommand{\phiqfi}{\ensuremath{\phi_{\mathrm{QFI}}}}
\newcommand{\phiecho}{\ensuremath{\phi_{\mathrm{echo}}}}
\newcommand{\cHeisQFI}{\ensuremath{c_\mathrm{QFI}}}
\newcommand{\cHeisSq}{\ensuremath{c_\mathrm{sq}}}
\newcommand{\Hxxz}{\ensuremath{H_\mathrm{XXZ}}}

\begin{document}

\preprint{APS/123-QED}

\title{Squeezing Towards the Heisenberg Limit with Locally Interacting Spins}

\author{Nazli~Ugur~Koyluoglu}
\affiliation{Department of Physics, Stanford University, Stanford, California 94305, USA}
\affiliation{Department of Physics, Harvard University, Cambridge, Massachusetts 02138, USA}
\affiliation{Harvard Quantum Initiative, Harvard University, Cambridge, Massachusetts 02138, USA}
\author{Shankari~V.~Rajagopal}
\affiliation{Department of Physics, Stanford University, Stanford, California 94305, USA}
\affiliation{Department of Physics, University of Michigan, Ann Arbor, Michigan, 48109, USA}
\author{Gabriel~L.~Moreau}
\affiliation{Department of Physics, Stanford University, Stanford, California 94305, USA}
\author{Jacob~A.~Hines}
\affiliation{Department of Physics, Stanford University, Stanford, California 94305, USA}
\affiliation{Department of Applied Physics, Stanford University, Stanford, California 94305, USA}
\author{Ognjen~Markovi\'{c}}
\affiliation{Department of Physics, Stanford University, Stanford, California 94305, USA}
\author{Monika~Schleier-Smith}
\affiliation{Department of Physics, Stanford University, Stanford, California 94305, USA}
\date{\today}

\begin{abstract}
We propose a robust approach to spin squeezing with local interactions that approaches the Heisenberg limit of phase sensitivity. To generate the requisite entanglement, we generalize the paradigmatic two-axis countertwisting Hamiltonian—akin to squeezing by parametric amplification—to systems with power-law interactions, incorporating a Heisenberg coupling that aids in spreading correlations and protects the collective spin coherence. The resulting time to approach the Heisenberg limit scales sublinearly with particle number in 2D dipolar and 3D van der Waals interacting systems. Our protocol is robust to disorder and density fluctuations, and can be implemented in near-term experiments with molecules, Rydberg atoms, and solid-state spins.
\end{abstract}

\maketitle

Entanglement is a powerful resource for enhancing precision measurements of forces, fields, accelerations, and time~\cite{pezze2018quantum,degen2017quantum,ye2024essay}. A broadly applicable form of entanglement is spin squeezing, in which nonclassical correlations suppress quantum projection noise~\cite{kitagawa1993squeezed,wineland1994squeezed}.  This noise suppression can enable enhanced phase resolution $\Delta \phi $\;$=$\;$ 1/\sqrt{GN}$ in a system of $N$ particles, where achieving a metrological gain $G$\;$>$\;$1$ requires $G$-particle entanglement~\cite{sorensen2001entanglement}.  Approaching the fundamental Heisenberg limit $G$\;$=$\;$N$ thus requires a fully entangled system.  To efficiently and scalably generate entanglement, seminal theory and experiments focused on all-to-all interactions~\cite{kitagawa1993squeezed,wineland1994squeezed,liu2011spin,leibfried2004toward, esteve2008squeezing, riedel2010atom, leroux2010implementation, lucke2011twin, hamley2012spin, berrada2013integrated, hosten2016quantum, bohnet2016quantum, pedrozo2020entanglement, greve2022entanglement, li2023improving}.  However, pioneering proposals for spin squeezing with local interactions~\cite{rey2008many,frerot2017entanglement,kwasigroch2017synchronization,kaubruegger2019variational,perlin2020spin, bilitewski2021dynamical,comparin2022robust,comparin2022scalable, block2024scalable} have led to recent experimental demonstrations with Rydberg atoms~\cite{hines2023spin, eckner2023realizing, bornet2023scalable}, trapped ions~\cite{franke2023quantum}, ultracold atoms in optical lattices~\cite{lee2024observation,douglas2024spin}, and solid state spins~\cite{wu2025spin}, as well as progress towards squeezing with dipolar molecules~\cite{christakis2023probing,holland2023ondemand,miller2024twoaxis}.

\begin{figure}[tb]
\includegraphics[width=\columnwidth]{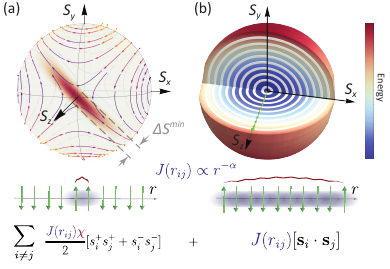}
\caption{\textbf{Gap-protected countertwisting} for squeezing with local interactions.  \textbf{(a)} The countertwisting terms squeeze the $\Smin=S^{\pi/4}$ quadrature of the collective spin (top) by generating pairwise correlated spin excitations (bottom). \textbf{(b)} Heisenberg interactions protect the collective spin coherence by energetically constraining the dynamics to the permutation symmetric subspace $S=N/2$ (top), or equivalently by allowing correlations to spread spatially (bottom).}
\label{fig:fig1}
\end{figure}

In engineering interactions for spin squeezing, two key considerations are the attainable metrological gain and the speed of achieving it.  By these metrics, an ideal route employs the countertwisting Hamiltonian~\cite{kitagawa1993squeezed} $\Hct \propto (S_+^2 + S_-^2)/S$ for $N$\;$=$\;$2S$ spin-1/2 particles forming a collective spin $\vec{S} $\;$=$\;$ \sum_{i=1}^N \vec{s}_i$. Here, global interactions generate pairwise correlated spin excitations.  This process, akin to parametric amplification, theoretically squeezes exponentially quickly [Fig.~\ref{fig:fig1}(a)], producing a Heisenberg scaling $G \propto N$ in a time $t $\;$\propto$\;$ \log N$.  A more widely realized approach of one-axis twisting~\cite{kitagawa1993squeezed} leverages collective Ising interactions, described by $\Htwist $\;$\propto$\;$ S_x^2/S$,  to squeeze by a factor $G \propto N^{2/3}$ in a time $t $\;$\propto$\;$ N^{1/3}$.  Extending either method to systems with short-range interactions generically may require either reducing the scaling of the metrological gain or extending the time~\cite{comparin2022scalable,block2024scalable}, due to speed limits on the propagation of correlations~\cite{lieb1972finite,tran2021lieb,tran2021optimal}.  

Even after allowing time for correlations to spread, local interactions do not typically produce correlations of the form required for strong squeezing. Squeezing consists of reducing the phase uncertainty $\Delta\phi $\;$=$\;$ \Delta \Smin/\abs{\avg{\vec{S}}}$, where $\Delta \Smin$ denotes the minimum quantum noise of spin components transverse to the mean $\avg{\vec{S}}$ [Fig.~\ref{fig:fig1}(a)]. Maximizing the signal $\abs{\avg{\vec{S}}}$ while reducing noise requires permutation-symmetric correlations. To preserve permutation symmetry even for local interactions, a powerful approach leverages Heisenberg interactions $\vec{s}_i\cdot\vec{s}_j$ to produce an energy gap between triplet and singlet states for each pair $(i,j)$ of spins, and correspondingly between manifolds of different total spin $S$ [Fig.~\ref{fig:fig1}(b)]~\cite{rey2008many,perlin2020spin}. The resulting spin-exchange processes allow quantum correlations to spread spatially, while conserving all moments of the collective spin. Gap protection can thus be seamlessly combined with additional interactions that squeeze the quantum noise.

A widely studied example is gap-protected one-axis twisting, which is realized by anisotropic XXZ Heisenberg models~\cite{perlin2020spin,frerot2017entanglement,bilitewski2021dynamical, comparin2022scalable,comparin2022robust,roscilde2023entangling,block2024scalable}. For globally interacting spins, an energy gap between manifolds of different total spin has been observed~\cite{norcia2018cavity} and applied to suppress inhomogeneous dephasing~\cite{davis2020protecting,niu2024many}.  For power-law interactions, the analogous gap protection suppresses interaction-induced dephasing to facilitate spin squeezing, as shown theoretically by Perlin et al.~\cite{perlin2020spin} and demonstrated in experiments with dipolar spin systems~\cite{bornet2023scalable,wu2025spin}. Recently, pioneering experiments have also engineered gap-protected countertwisting and probed the resulting mean-field dynamics in both globally~\cite{luo2025hamiltonian} and locally~\cite{miller2024twoaxis,gao2025signal} interacting systems.  Despite the theoretical advantages in the all-to-all case, the speed and scaling of squeezing by local countertwisting has yet to be investigated.

In this paper, we analyze the squeezing dynamics of the gap-protected countertwisting Hamiltonian 
\begin{equation}\label{eq:Hctg}
\Hctg = \sum_{i\neq j} J(r_{ij}) \left[\vec{s}_i\cdot \vec{s}_j  + \frac{\ginv}{2}\left(s^+_i s^+_j + s^-_i s^-_j\right)\right]
\end{equation}
with power-law interactions $J $\;$\propto$\;$ r^{-\alpha}$. For sufficiently slow squeezing, we recover the Heisenberg scaling $G$\;$\propto$\;$ N$ of the globally interacting model. Applying spin-wave theory to predict the allowable rate of squeezing with $N $\;$=$\;$ L^d$ spins in $d$ spatial dimensions, we find a time $t_\mathrm{H} $\;$\propto$\;$ L^2\log(N)$ to reach Heisenberg scaling in the limit of short-range interactions, which improves to a nearly linear scaling $t_\mathrm{H} $\;$\propto$\;$ L\log(N)$ with length $L$ for dipolar interactions in $d$\;$=$\;$2$ dimensions. These predictions are corroborated by numerical simulations, which further establish the robustness of the approach to disorder and density fluctuations.  Our results lay groundwork for elucidating universal principles governing the speed and scaling of squeezing with local interactions.

Our focus on the gap-protected countertwisting Hamiltonian is motivated by a spin-wave analysis of the early-time squeezing dynamics of a generic XYZ model $H_\mathrm{XYZ} $\;$=$\;$ \sum_{i,j,\mu} J^\mu(\vec{r}_{ij}) s^\mu_i s^\mu_j$, where $\mu$\;$\in$\;$\left\{x,y,z\right\}$ indexes the spin components. We consider $N$ spins arranged on a lattice, enabling an analysis in momentum space, and initialized in a spin-polarized state along $\uvec{z}$. The early-time dynamics can then be examined in the Holstein-Primakoff approximation, where spin-wave operators $S^+_\k$ are approximated as bosonic creation operators $a^\dagger_{-\k} $\;$\approx$\;$ S^+_\k $\;$\equiv$\;$ \frac{1}{\sqrt{N}}\sum_{j} e^{-i \vec{k}\cdot\vec{r}_j}\vec{s}^+_j$.  In this approximation, $H_\mathrm{XYZ}$ reduces to the form
\begin{equation}\label{eq:Hhp}
H_\mathrm{XYZ} \approx \sum_k \left[\chi_\k \left(\frac{a^\dagger_{-\k} a^\dagger_{\k} + a_\k a_{-\k}}{2}\right) + \omega_\k a^\dagger_\k a_{\k} \right],
\end{equation}
where $\omega_\k$ quantifies the energy cost of creating a spin wave of momentum $\k$ and $\chi_\k$ parameterizes the rate of squeezing in momentum modes $\pm\k$ by exciting opposite-momentum spin-wave pairs. In terms of momentum-space couplings $\tJ_\k^\mu $\;$\equiv$\;$ \sum_\vec{r} e^{-i \vec{k}\cdot \vec{r}} J^\mu(\vec{r})$, we have $\omega_\k $\;$=$\;$ \left(\tJ^x_\k+\tJ^y_\k\right)/2 - \tJ^z_\vec{0}$ and $\chi_\k $\;$=$\;$ \left(\tJ^x_\k - \tJ^y_\k\right)/2$.

For applications in sensing a global field, we aim to maximize squeezing of the $\k $\;$=$\;$ \vec{0}$ mode. We should thus set $\omega_\vec{0}$\;$=$\;$0$ to enable resonant parametric amplification of zero-momentum spin waves, while suppressing excitation of higher-momentum modes by setting a large ratio $\omega_\k/\chi_\k $\;$\gg$\;$ 1$ for $\k $\;$\neq$\;$ 0$.  The gap-protected countertwisting Hamiltonian (Eq.~\ref{eq:Hctg} and Fig.~\ref{fig:fig1}) achieves precisely these conditions.  Specifically, the couplings $\tJ_\k^\mu $\;$=$\;$ J^\mu f_\k$ can be factorized into the spin anisotropy $J^{x,y,z} $\;$=$\;$ (1+\ginv,\, 1-\ginv,\, 1)$ and the momentum-dependence $f_\k$, where we fix $f_\vec{0} $\;$=$\;$ 1$ so that the energy scales extensively.  The rate of global squeezing is then set by $\chi_\vec{0} $\;$=$\;$ \chi$, while excitation of spin-waves at nonzero momentum is suppressed by the ratio $\omega_\k/\chi_\k $\;$=$\;$ \left(1-1/f_\k\right)/\chi$ for sufficiently slow global squeezing rate $\chi$.

The requirement of keeping the countertwisting rate $\chi$ small is mitigated by the squeezing proceeding exponentially fast due to parametric resonance. In particular, diagonalizing the Hamiltonian in Eq.~\ref{eq:Hhp} by a Bogoliubov transformation yields eigenvalues $\lambda_\k $\;$=$\;$ \pm \sqrt{\omega_\k^2 -  \chi_\k^2}$.  The resulting stability condition $\abs{\omega_\k/\chi_\k} $\;$>$\;$ 1$ confirms that the $\k=\0$ mode, with $\omega_\vec{0} = 0$, is always unstable. Specifically, the early-time evolution of collective spin operators $S^\theta $\;$=$\;$ \cos\theta S^x_{\k=\0} + \sin\theta S^y_{\k=\0}$ in the Heisenberg picture is given by $dS^{\pm \pi/4}/dt $\;$\approx$\;$ \mp\chi S^{\pm \pi/4}$.  Thus, the noise $\Delta S^{\pi/4}$\;$\equiv$\;$ \Delta S^\mathrm{min}$ in the squeezed quadrature shrinks exponentially, approaching the Heisenberg limit $\Delta S^\mathrm{min} $\;$\sim$\;$ 1$ in a time $\chi t $\;$\sim$\;$ \log N$~\cite{kitagawa1993squeezed}.  Determining the scaling of the absolute time $t$ with $N$ requires additionally accounting for any $N$-dependence of the squeezing rate $\chi$.

To estimate how quickly one can squeeze while constraining the dynamics to the $\vec{k}$\;$=$\;$\0$ mode, we enforce the stability condition $\abs{\omega_\k/\chi_\k} $\;$>$\;$ 1$ for all nonzero wavenumbers. For example, for power-law interactions $J_{ij} $\;$=$\;$ r_{ij}^{-\alpha}$ in $d<\alpha$ spatial dimensions, the dispersion relation $\omega_\k=f_\k-1$ takes the form $\omega_\k $\;$\propto$\;$  k^{\gamma(\alpha,d)}$ for $k$\;$\ll$\;$ 1$, where $\gamma(\alpha,d) $\;$=$\;$ \min(2,\alpha-d)$~\cite{frerot2017entanglement}.  Defining $\kmin $\;$\propto$\;$ N^{-1/d}$ as the smallest nonzero wavenumber in a $d$-dimensional lattice, we predict that the squeezing rate should be limited to
\begin{equation}\label{eq:gscaling}
\chi < 1-1/f_{\kmin} \propto N^{-\min(2,\alpha-d)/d}.
\end{equation}
The dependence on $N$ is consistent with requiring time for correlations to spread across the entire system in order to recover the Heisenberg scaling of the all-to-all model.

\begin{figure}[tb]
\includegraphics[width=\columnwidth]{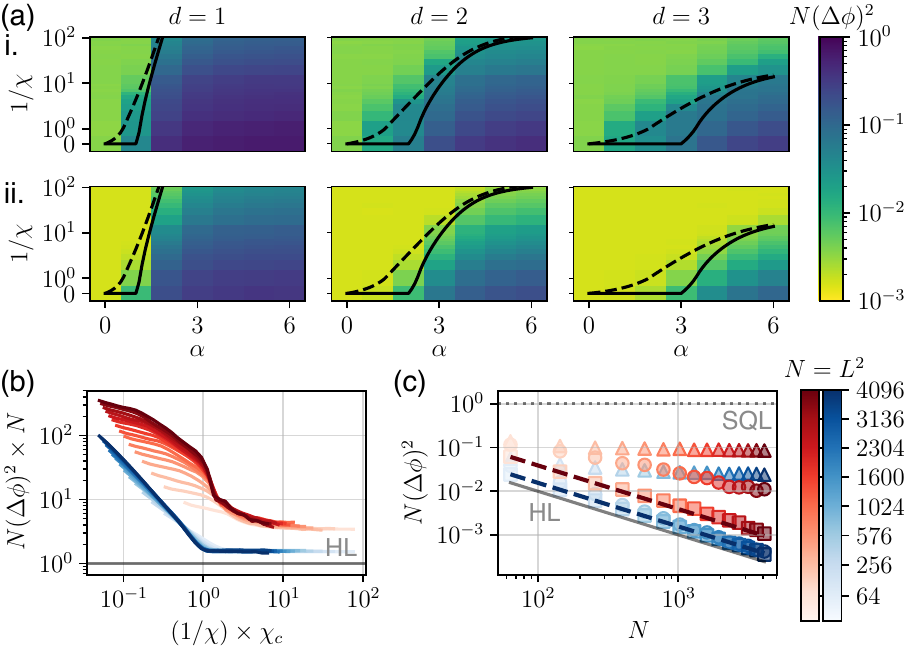}
\caption{\textbf{Enhanced sensitivity by gap-protected countertwisting.} \textbf{(a)} Sensitivities
(i) $N(\Delta\phisq)^2$  and (ii) $N(\Delta\phiqfi)^2$ vs inverse squeezing rate $\chi^{-1}$ and power-law exponent $\alpha$ for $N=10^3$ spins in $d$ dimensions with open boundary conditions. Black curves indicate spin-wave prediction of critical value $\chi_c^{-1}$ for periodic (solid) and open (dashed) boundary conditions. \textbf{(b-c)} Sensitivity $N(\Delta\phi_\mathrm{sq/QFI})^2$ (red/blue) in $(d,\alpha)$\;$=$\;$(2,3)$, calculated with periodic boundary conditions for varying $N $\;$\le$\;$ 4096$.  (b) Sensitivity relative to the Heisenberg limit (solid gray line) vs gap protection $\chi^{-1}$ rescaled by critical value $\chi_c^{-1}$ for each $N$.  (c) Sensitivity relative to standard quantum limit (dotted gray line) vs system size $N$ at weak gap protection $\chi = 1$ (triangles), at criticality $\chi $\;$=$\;$ \chi_c$ (circles), and at strong gap protection $\chi $\;$=$\;$ 0.2 \chi_c$ (squares). Dashed curves show optimal squeezing (red) and QFI (blue) for all-to-all interactions ($\alpha $\;$=$\;$ 0$).}
\label{fig:fig2}
\end{figure}

To verify the predictions of the spin-wave analysis, we numerically simulate gap-protected countertwisting using the discrete truncated Wigner approximation (DTWA), a semiclassical phase-space method for investigating early-time quantum many-body dynamics~\cite{schachenmayer2015many,SM}. We compute the time evolution of the Wineland squeezing parameter 
$N(\Delta\phisq)^2$, which quantifies the enhancement of angular resolution $\Delta \phisq$\;$=$\;$\Delta \Smin/\langle S_z \rangle$ compared to the standard quantum limit $\Delta\phi_{\text{SQL}}$\;$=$\;$1/\sqrt{N}$~\cite{wineland1992spin}. The squeezing reaches an optimum at a time $\chi t_\mathrm{opt} $\;$\lesssim$\;$ \log(N)$~\cite{SM}, and we first focus on examining the squeezing at this optimal time.

We plot the dependence of the optimal squeezing on the rate $\chi$ in Fig.~\ref{fig:fig2}(a)(i), where we consider power law exponents in the range $0$\;$\leq$\;$ \alpha $\;$\leq$\;$ 6$ and spatial dimensionalities $d$\;$=$\;$1,2,3$ at fixed particle number $N$\;$=$\;$10^3$. For a sufficiently slow countertwisting rate $\chi$, the optimal squeezing comes close to the Heisenberg limit, even with local interactions. In other words, increasing the strength of gap protection $\chi^{-1}$ mitigates the effect of locality and extends the regime of favorable squeezing towards larger power-law exponent $\alpha$. Promising regimes for implementations are dipolar interactions ($\alpha $\;$=$\;$ 3$) in two dimensions~\cite{wu2025spin} and van der Waals interactions ($\alpha $\;$=$\;$ 6$) in three dimensions~\cite{hines2023spin}. In both cases, a gap protection strength $\chi^{-1}$\;$\sim$\;$ 10^2$ allows the squeezing parameter to approach the value $N(\Delta\phisq)^2 $\;$=$\;$ \cHeisSq/N $ attained by the ideal all-to-all countertwisting ($\alpha $\;$=$\;$ 0$), where $\cHeisSq\approx 3.9$.

In approaching the Heisenberg limit, the quantum state necessarily becomes non-Gaussian and the squeezing no longer fully captures the metrological gain. Nevertheless, the large antisqueezing $\Delta\Smax$ in a pure state implies high sensitivity to rotations about the $\Smax$ axis.  The quantum Fisher information (QFI) $F$\;$=$\;$4(\Delta \Smax)^2$ quantifies this enhanced sensitivity $N(\Delta\phiqfi)^2 $\;$=$\;$ N/F$ ~\cite{braunstein1994statistical,giovannetti2011advances}, which we plot in Fig.~\ref{fig:fig2}(a)(ii). Experimentally, an echo protocol --- in which the sign of interactions is reversed prior to state readout~\cite{davis2016approaching,macri2016loschmidt,anders2018phase} --- helps to access this enhanced sensitivity~\cite{SM}.

Both measures of sensitivity $N(\Delta\phi^2)$ in Fig.~\ref{fig:fig2}(a) show a qualitatively similar dependence on the power-law exponent $\alpha$, dimension $d$, and strength of gap protection $\chi^{-1}$. The results are consistent with the critical value $\chi_c $\;$=$\;$ 1-1/f_{\kmin}$ predicted by spin-wave theory (Eq.~\ref{eq:gscaling}), shown as black solid/dashed curves calculated with periodic/open boundary conditions~\cite{SM}. While gap protection is essential for short-range interactions $\alpha $\;$>$\;$ d$, it is also beneficial for $\alpha $\;$\lesssim$\;$ d$ in the realistic case of open boundary conditions.

To verify that the sensitivity approaches a Heisenberg scaling above the critical gap protection strength $\chi_c^{-1}$, we examine a scaling collapse across system sizes for a representative case $(d, \alpha) $\;$=$\;$ (2, 3)$ in Fig.~\ref{fig:fig2}(b).  The QFI (blue curves), plotted relative to the Heisenberg limit, features a universal improvement in sensitivity with increasing gap protection $\chi^{-1}$ up to $\chi_c/\chi$\;$=$\;$ 1$, followed by saturation a constant factor $\cHeisQFI \approx 1.56$ away from the Heisenberg limit at $\chi$\;$\lesssim$\;$ \chi_c$.  The squeezing (red curves) exhibits $N$-dependence up to $\chi_c^{-1}$, followed by a collapsed convergence towards a factor $\cHeisSq$ away from the Heisenberg limit at $\chi\lesssim \chi_c$. In Fig.~\ref{fig:fig2}(c), we further analyze the system-size scaling of these quantities in various regimes of gap protection, parameterized by squeezing rates $\chi $\;$=$\;$ 1$ (triangles), $\chi $\;$=$\;$ \chi_c$ (circles), and $\chi/\chi_c $\;$=$\;$ 0.2$ (squares). Neither the squeezing (red) nor the QFI (blue) scales with system size at fixed $\chi=1$. However, the Heisenberg scaling can be recovered by setting $\chi$\;$=$\;$\chi_c$ for the QFI, and $\chi $\;$\ll$\;$ \chi_c$ for squeezing.

\begin{figure}[tb!]
\includegraphics[width=\columnwidth]{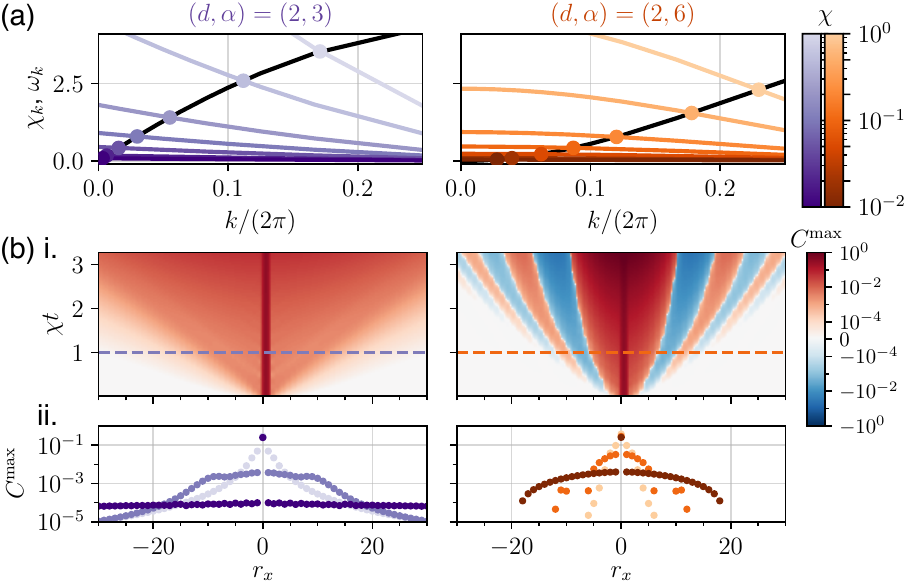}
\caption{\textbf{Mechanism for achieving collective behavior}, visualized for 2D systems with $\alpha = 3, 6$ (left, right). \textbf{(a) Momentum-space picture.} Spin-wave dispersion $\omega_k$ (black curves) and parametric coupling $\chi_k$ (colored curves, dark to light for increasing $\chi$). Markers at $\omega_{k_c} = \chi_{k_c}$ indicate critical system size $N_c$\;$\propto$\;$ k_c^{-d}$ below which the system exhibits Heisenberg scaling for the given strength of gap protection $\chi^{-1}$. \textbf{(b)~Real-space picture.} Linear spin-wave theory for $N=64^2 $\;$=$\;$ 4096$ spins with periodic boundary conditions, showing (i) correlations $\Cmax(r_x) $\;$=$\;$ \Cmax_{i,i+r_x}$ vs. distance $r_{x}$ and time $t$ at $\chi$\;$=$\;$0.1$ and (ii) cuts of $\Cmax(r_x)$ at fixed time $\chi t $\;$=$\;$ 1$, for $\chi=1,\;0.1,\;0.01$ (light to dark).}
\label{fig:fig3}
\end{figure}

To understand the metrological gain attainable when squeezing faster than the critical rate $\chi_c$, we examine the spatial structure of the correlations generated by $\Hctg$. We are motivated by expressing the QFI as a sum over pairwise correlations in the antisqueezed quadrature, $F $\;$=$\;$ 4\sum_{ij} \Cmax_{ij}$, where $C^\mu_{ij} $\;$\equiv$\;$ \avg{s_i^\mu s_{j}^\mu} - \avg{s_i^\mu}\avg{s_{j}^\mu} \leq 1/4$.  Achieving a metrological gain $G=1/[N(\Delta\phi)^2]= F/N$ thus requires that correlations extend over a volume of $G$ sites.  The early-time growth of correlations is governed by the spin-wave dispersion $\omega_k$ and parametric coupling $\chi_k$, which we plot in Fig.~\ref{fig:fig3}(a) for $\alpha $\;$=$\;$ 3$ (left) and $\alpha $\;$=$\;$ 6$ (right) in two dimensions. The stability condition $\abs{\chi_k / \omega_k} $\;$<$\;$ 1$ sets a wavenumber $k_c(\chi)$ below which parametric amplification occurs, marked by circles showing $k_c$ increasing with squeezing rate $\chi$. The resulting correlations in real space, plotted in Fig.~\ref{fig:fig3}(b) using linear spin-wave theory (Eq.~\ref{eq:Hhp}), exhibit a plateau of size $L_c $\;$\sim$\;$ 2\pi / k_c$ that shrinks with increasing squeezing rate.

We interpret the plateau size $L_c$ as the length-scale over which correlations spread uniformly during the time-scale $\chi^{-1}$ for squeezing.  Fig.~\ref{fig:fig3}(b)(ii) shows the correlations for different squeezing rates at early time $\chi t $\;$=$\;$ 1$, where the spin-wave theory applies~\cite{SM}.  The correlation length $L_c$ is then governed by the exponent $\gamma(\alpha, d)$ in the dispersion $\omega_k $\;$\propto$\;$ k^\gamma$, with the stability condition $k_c^\gamma $\;$\approx$\;$ \chi$ yielding $L_c $\;$\propto$\;$ \chi^{-1/\gamma}$.  The growth of the plateau with increasing gap protection $\chi^{-1}$ in Fig.~\ref{fig:fig3} is thus faster for the linear dispersion ($\gamma $\;$=$\;$ 1$ for $\alpha $\;$=$\;$ 3$; left) than for the quadratic dispersion ($\gamma $\;$=$\;$ 2$ for $\alpha $\;$=$\;$ 6$; right).  From the correlation length, we predict a scaling $G $\;$\propto$\;$ L_c^d $\;$\propto$\;$ \chi^{-d/\gamma}$ in $d$ spatial dimensions.  The scaling in Fig.~\ref{fig:fig2}(b) is consistent with this prediction up to finite-size effects~\cite{SM}.

From the dependence of metrological gain on squeezing rate, we find the critical rate $\chi_c $\;$\propto$\;$ N^{-\gamma/d}$ for reaching Heisenberg scaling $G $\;$\propto$\;$ N$.  This rate is slowest in low-dimensional systems with local interactions $\alpha $\;$\geq$\;$ d+2$ that yield a quadratic dispersion ($\gamma $\;$=$\;$ 2$), as we saw in Fig.~\ref{fig:fig2}(a) (black curves showing $\chi_c^{-1}$).  The rate $\chi_c$ sets a time $t_H $\;$\sim$\;$ \log N / \chi_c $\;$\sim$\;$ N^{\gamma/d}\log N$ to reach a Heisenberg scaling in metrological gain.  If practical constraints preclude reaching the time $t_H$, a reduced metrological gain $G$ can be achieved by squeezing at a faster rate $\chi $\;$\sim$\;$ G^{-\gamma/d}$ for time $t $\;$\sim$\;$ G^{\gamma/d}\log G$.  For example, realistic cases $(d,\alpha) $\;$=$\;$ (2,3)$ and $(3,6)$ permit sublinear scalings of time with metrological gain due to the comparatively small exponents $\gamma/d $\;$=$\;$ 1/2$ and $\gamma/d $\;$=$\;$ 2/3$.  By contrast, short-range interactions ($\alpha $\;$\ge$\;$ d+2$) in $d=1$ dimension require a significantly longer time $t $\;$\sim$\;$ G^2 \log G$.

\begin{figure}[tb]
\includegraphics[width=\columnwidth]{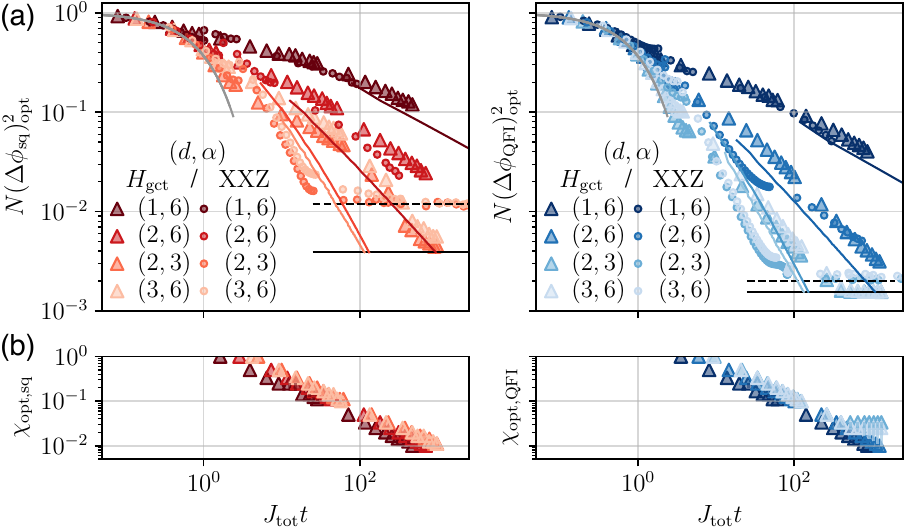}
\caption{\textbf{Time-optimal protocols.} \textbf{(a)} DTWA simulations of sensitivity $N(\Delta\phi_\mathrm{sq/QFI})^2$ (left/right) at optimal squeezing rate $\chi$ vs total interaction time $\Jtot t$, comparing $\Hctg$ (triangles) with $\Hxxz$ (circles)
for $N=1000$ spins in various configurations $(d, \alpha)$. We overlay analytical predictions (solid) for optimal gain under $\Hctg$ at short (grey) and intermediate (red/blue) times $\Jtot t$~\cite{SM}, and saturation values (black) predicted from all-to-all models $\Hct$ (solid) and $\Htwist$ (dashed). \textbf{(b)} Optimal squeezing rate $\chi_\mathrm{opt}$ vs $\Jtot t$ for $\Hctg$.}
\label{fig:fig4}
\end{figure}

To corroborate the predictions of the spin-wave analysis, Fig.~\ref{fig:fig4}(a) shows DTWA simulations of the sensitivity $N(\Delta\phi)^2 $\;$=$\;$ 1/G$ vs evolution time for several choices $(\alpha, d)$ of power-law exponent and dimensionality. We assume that $\Hctg$ is generated by Floquet engineering~\cite{liu2011spin,miller2024twoaxis}, so that the sum $J_\mathrm{tot} $\;$=$\;$ \sum_\mu \abs{J^\mu} $\;$=$\;$ \abs{1+\chi} + \abs{1-\chi} + 1$ of coupling strengths over Cartesian components $\mu$ determines the total evolution time $\Jtot t$.  In the limit $\chi$\;$\ll$\;$ 1$ required for large metrological gain, $\Jtot t $\;$\rightarrow$\;$ 3t$.  We plot the squeezing (red triangles) and QFI (blue triangles) optimized over the choice of squeezing rate $\chi$ for each evolution time $\Jtot t$.

The dependence of metrological gain on total interaction time $\Jtot t$ is well described by a simplified model (solid curves in Fig.~\ref{fig:fig4}) of the optimal squeezing rate $\chi$ in three regimes~\cite{SM}. At small $\Jtot t$, where the dominant limitation is the short evolution time, fast squeezing $\chi \to \infty$ is optimal. With increasing $\Jtot t$, as the dominant limitation becomes the short interaction range, the rate $\chi$ should be reduced to extend the correlation length, resulting in a predicted sensitivity $N(\Delta\phi)^2 \propto \chi^{d/\gamma}$ that agrees well with the simulated QFI and also qualitatively captures the squeezing.  Ultimately, decreasing the squeezing rate below the critical value $\chi_c$ provides diminishing returns, as the sensitivity saturates to the ideal, near-Heisenberg-limited values $N(\Delta\phi_{\mathrm{sq}/\mathrm{QFI}})^2 $\;$=$\;$ c_{\mathrm{sq}/\mathrm{QFI}}/N$ of the collective model $\Hct$. At the longest simulated times, we observe this saturation for favorable combinations $(d,\alpha) $\;$=$\;$ (2,3)$ and $(d,\alpha) $\;$=$\;$ (3,6)$ of dimension and power-law exponent.

For comparison, we also plot the optimal metrological gain vs total interaction time for gap-protected one-axis twisting (circles in Fig.~\ref{fig:fig4}), generated by an XXZ model $\Hxxz$ in a rotated basis $J^{x,y,z}$\;$=$\;$(1-\chi,1,1)$ chosen to squeeze in the $xy$-plane. While the short-time behavior closely resembles that under $\Hctg$, the long-time saturation is limited by the ideal squeezing and QFI values of the all-to-all model $\Htwist$, allowing $\Hctg$ to provide stronger squeezing for times $t$\;$\gtrsim$\;$ t_H$. At intermediate times, and for sufficiently long-ranged interactions, $\Hxxz$ outperforms $\Hctg$, exhibiting a sudden improvement at a finite anisotropy value $1-\chi $\;$\sim$\;$ \mathcal{O}(1)$ thanks to a phase transition into finite-temperature easy-plane ferromagnetic order~\cite{perlin2020spin,block2024scalable}.  Absent this phase transition, e.g., in the short-ranged setting $(d, \alpha) $\;$=$\;$ (1,6)$, the squeezing performance of the gap-protected twisting ($\Hxxz$) and countertwisting models is nearly identical before saturation, suggesting that the analysis of $\Hctg$ captures universal aspects of squeezing with local interactions.

\begin{figure}[tb]
\includegraphics[width=\columnwidth]{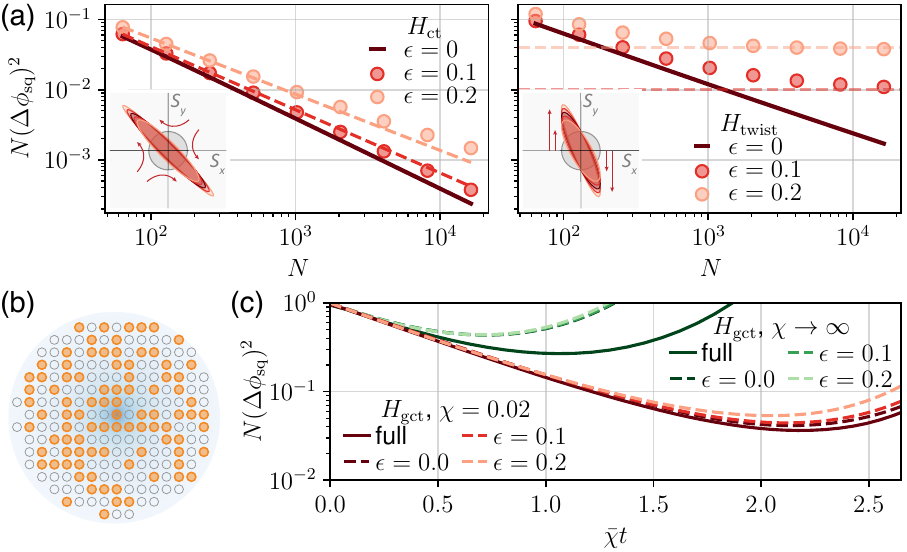}
\caption{\textbf{Robustness of squeezing.}  (a) Comparison of squeezing under $\Hct$ (left) and $\Htwist$ (right) with rms fractional fluctuation $\epsilon$ in the strength of global interactions ($\alpha$\;$=$\;$0$).  Markers show $\epsilon $\;$=$\;$ 0.1, 0.2$ (dark to light), while solid curve shows the ideal squeezing for $\epsilon $\;$=$\;$ 0$.  Dashed curves show expected scaling $N(\Delta\phisq)^2 $\;$\propto$\;$ N^{1-\epsilon}$ for $\Hct$ and saturation value $N(\Delta\phisq)^2 $\;$\approx$\;$ \epsilon^2$ for $\Htwist$.  (b-c) Effects of disorder and density fluctuations for $(d,\alpha) $\;$=$\;$ (2,3)$.  We consider $n$ atoms randomly distributed over $N$\;$=$\;$200$ lattice sites, as shown in (b), with fluctuating filling fraction $n/N $\;$=$\;$ 0.5 (1 \pm \epsilon)$.  Dashed curves show squeezing under $\Hctg$ for $\epsilon $\;$=$\;$ 0, 0.1, 0.2$ (dark to light) with $\chi $\;$=$\;$ 0.02$ (red) and in the limit $\chi $\;$\rightarrow$\;$ \infty$ of no gap protection (green).  Solid curves show dynamics of a uniformly filled array of $N$\;$=$\;$100$ sites.
}
\label{fig:fig5}
\end{figure}

Despite similarities, an advantage of squeezing with $\Hctg$ over the XXZ model is its robustness to density fluctuations and disorder.  We expect countertwisting to be advantageous because the squeezed quadrature has a fixed orientation, in contrast to one-axis twisting where the orientation depends on the strength of squeezing [Fig.~\ref{fig:fig5}(a) insets].  To verify this intuition, we first consider squeezing with a fluctuating collective interaction strength under the all-to-all models $\Hct$ and $\Htwist$.  For small fractional fluctuations $\epsilon$, the optimal squeezing under $\Hct$ [Fig.~\ref{fig:fig5}(a), left] retains a near-Heisenberg scaling $N(\Delta\phisq)^2 $\;$\propto$\;$ N^{-(1-\epsilon)}$, whereas for $\Htwist$ the squeezing saturates to a constant value $N(\Delta\phisq)^2 \sim \epsilon^2$ [Fig.~\ref{fig:fig5}(a), right].  These scalings (dashed lines) are predicted from the geometry of the squeezing dynamics~\cite{SM} and borne out by exact calculations (markers).

Since gap protection can recover the all-to-all behavior of $\Hct$, the advantage of countertwisting can persist even for local interactions.  In this case, randomly filling a fluctuating fraction $n/N$\;$=$\;$0.5 (1\pm\epsilon)$ of the lattice sites [Fig.~\ref{fig:fig5}(b)] produces fractional fluctuations $\epsilon$ in mean-field interaction strength, as we simulate in Fig.~\ref{fig:fig5}(c) for $n $\;$\approx$\;$ 100$ atoms in a 2D lattice with $\alpha $\;$=$\;$ 3$.  For reference, the green curves show the limit $\chi$\;$\rightarrow$\;$\infty$ of no gap protection, where even the squeezing at unity filling (solid curve) is limited by the locality of interactions.  Introducing positional disorder significantly degrades the squeezing, regardless of fluctuations $\epsilon$, because different clusters of spins squeeze at different rates set by the local density. By contrast, for small $\chi$\;$=$\;$0.02$ (red curves), the dynamics becomes collective, so that the effect of the random filling is only a slight reduction in metrological gain with increasing density fluctuations $\epsilon$. 

Gap-protected countertwisting may be realized on a variety of experimental platforms. These include dipolar spin systems ($\alpha$~=~3), such as cold molecules~\cite{holland2023ondemand, miller2024twoaxis}, Rydberg atoms~\cite{bornet2023scalable}, and ensembles of nitrogen-vacancy centers in diamond~\cite{gao2025signal,wu2025spin}, as well as systems with van der Waals interactions ($\alpha$~=~6), such as Rydberg-dressed atoms~\cite{hines2023spin, eckner2023realizing, young2023enhancing,bouchoule2002spin,glaetzle2015designing,steinert2023spatially}.  The countertwisting term can be generated either by Floquet engineering~\cite{liu2011spin,choi2020robust,geier2021floquet,miller2024twoaxis,gao2025signal,SM} from native XXZ or Ising interactions, or by inducing pairwise excitations via bichromatic driving~\cite{bouchoule2002spin,glaetzle2015designing,borregaard2017one,steinert2023spatially,luo2025hamiltonian}. In regimes where the squeezing does not capture the full benefit of the QFI, the metrological gain can be enhanced by an echo protocol that involves reversing the sign of the interactions~\cite{davis2016approaching,macri2016loschmidt,anders2018phase,colombo2022time,SM}.  Such echoes have been demonstrated with dipolar molecules~\cite{miller2024twoaxis} and NV centers~\cite{gao2025signal}, and could be extended to Rydberg-dressed atoms~\cite{davis2016approaching,potirniche2017floquet}. 

We have introduced a robust method of spin squeezing towards the Heisenberg limit, combining local countertwisting to generate pairwise excitations with local Heisenberg interactions to spread these correlations.  As the countertwisting terms perturbatively break the SU(2) symmetry that protects the collective spin, we derived requirements on how slow the countertwisting must proceed to achieve a target metrological gain.  The metrological gain reaches a Heisenberg scaling in time $t_H$\;$\sim$\;$ N^{1/2}\log{N}$ for 2D dipolar interactions and $t_H$\;$\sim$\;$ N^{2/d}\log{N}$ for short-range interactions in $d$ dimensions. An open question is whether these timescales can be further improved to saturate known Lieb-Robinson bounds~\cite{tran2021lieb,tran2021optimal}. The natural decomposition of our model into distinct ``squeezer'' (countertwisting) and ``spreader'' (Heisenberg) terms opens prospects for optimizing the squeezing via spatiotemporal control, guided by efficient simulations via our linear spin-wave framework. 

\begin{acknowledgments}
This work was supported by the Army Research Office under grant No. W911NF-20-1-0136.  We additionally acknowledge support from the Air Force Office of Scientific Research under grant No. FA9550-20-1-0059 (J.~A.~H., M.~S.-S.), the Department of Energy Q-NEXT Quantum Information Science Research Center (S.~V.~R.), the Office of Naval Research under grant No. N00014-17-1-2279 (O.~M.), the National Science Foundation Graduate Research Fellowship Program (G.~L.~M.), and the National Defense Science and Engineering Graduate Research Fellowship program (J.~A.~H).  We thank V. Khemani, N.~T. Leitao, L.~S. Martin, H. Gao, M.~Block, B.~Ye for valuable discussions, and T. Nebabu for contributions to the numerical simulation code.
\end{acknowledgments}

\bibliography{ctgap}

\clearpage
\onecolumngrid

\input{supplement_arxiv}

\end{document}

%% file: supplement_arxiv.tex
\setcounter{equation}{0}
\setcounter{page}{1}
\setcounter{figure}{0}
\renewcommand{\thepage}{S\arabic{page}} 
\renewcommand{\theequation}{S\arabic{equation}}
\renewcommand{\thefigure}{S\arabic{figure}}
\renewcommand{\theHfigure}{S\arabic{figure}}
\renewcommand{\thetable}{S\arabic{table}}
\renewcommand{\thesection}{\Roman{section}}
\renewcommand{\thesubsection}{\arabic{subsection}}
\newcommand{\zetamin}{\ensuremath{\zeta_\mathrm{min}}}
\newcommand{\zetamax}{\ensuremath{\zeta_\mathrm{max}}}
\newcommand{\Htfi}{\ensuremath{H_\mathrm{tfi}}}
\setcounter{secnumdepth}{4}

\begin{center}
\textbf{Squeezing Towards the Heisenberg Limit with Locally Interacting Spins:}\\
\textbf{Supplemental Material}\\ \vspace{10pt}
\end{center}
\maketitle

In this supplement, we present additional information about numerical and analytical methods employed in the main text and provide supporting analyses. In Sec.~\ref{sec:supp_dtwa}, we benchmark the semiclassical simulations against results from exact diagonalization in computationally tractable regimes.  Section~\ref{sec:supp_spin_wave} expands on the spin-wave theory for analyzing early-time squeezing dynamics, and validates its predictions against exact calculations of the Hamiltonian spectrum.  In Sec.~\ref{sec:supp_finitesize}, we elaborate on system-size scalings of the optimal sensitivity and the time to reach it. In Sec.~\ref{sec:echo_protocol}, we analyze an echo protocol for enhancement beyond squeezing. We expand on our analyses of time-optimized gap protection in Sec.~\ref{sec:supp_optimization}, and on robustness to system fluctuations in Sec.~\ref{sec:supp_robustness}. Finally, we discuss trotterized implementations of gap-protected countertwisting in Sec.~\ref{sec:supp_floquet}.

\section{DTWA simulations} \label{sec:supp_dtwa}

We use the discrete truncated Wigner approximation (DTWA), which is a semiclassical phase-space method for simulating quantum many-body dynamics of large ensembles of spin-1/2 particles~\cite{schachenmayer2015many}, and is widely used to compute metrological properties of time-evolved systems~\cite{perlin2020spin,block2024scalable}. In particular, we evaluate moments of collective spin observables statistically from Monte Carlo-sampled trajectories time-evolved according to classical equations of motion, and use these to compute metrics such as squeezing and quantum Fisher information (QFI).
We report observable expectation values through bootstrapping, i.e. by averaging over estimates from resampled subsets of simulated trajectories. We find that the statistical errors, computed as the standard error of the mean across resamples, are smaller than the markers in our DTWA simulations, including Fig.~\ref{fig:supp1} shown below. For results presented in Figs.~\ref{fig:fig2}(a),~\ref{fig:fig4},~\ref{fig:fig5} of the main text and in Figs.~\ref{fig:supp1},~\ref{fig:supp5},~\ref{fig:supp6},~\ref{fig:supp7} of the Supplemental Material, we use $n_t = 10^4$ trajectories and perform bootstrapping by resampling $10^2$ estimates with $10^3$ trajectories each. For Fig.~\ref{fig:fig2}(b-c) of the main text and in Figs.~\ref{fig:supp3},~\ref{fig:supp4} of the Supplemental Material, we instead use $n_t = 2000$ trajectories sampled once. 

\begin{figure}[h]
\includegraphics[width=0.75\columnwidth]{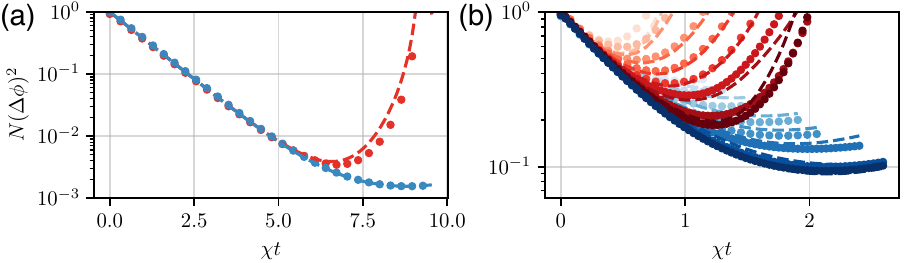}
\caption{\textbf{Benchmarking DTWA simulations.} \textbf{(a)} Sensitivity obtained from squeezing parameter $N(\Delta \phisq^2$ (red) and QFI $N(\Delta \phiqfi)^2$ (blue) vs. time $\chi t$, for $N=1000$ all-to-all interacting spins evolved under the canonical two-axis countertwisting Hamiltonian $\Hct$, comparing DTWA (markers) and exact diagonalization (dashed curves). \textbf{(b)} Sensitivity $N(\Delta \phi_{\mathrm{sq/QFI}})^2$ (red/blue) vs. $\chi t$ for $N=16$ spins in $(d,\alpha)=(1,6)$, evolved under $\Hctg$ with gap protection strength $\chi^{-1}=0,1,2,5,10,20,50,100$ (light to dark), comparing DTWA (markers) and exact diagonalization (dashed curves).}
\label{fig:supp1}
\end{figure}

To verify that the DTWA simulations accurately predict the optimal sensitivity and evolution time, we benchmark the simulations against exact diagonalization in numerically tractable regimes in Fig.~\ref{fig:supp1}. The markers in Fig.~\ref{fig:supp1}(a) show DTWA simulations of squeezing (red) and QFI (blue) for all-to-all countertwisting $\Hct $\;$=$\;$ (S_+^2 + S_-^2)/S$ with $N = 10^3$ spins.  The DTWA simulations are in good agreement with the evolution calculated by exact diagonalization (dashed curves), capturing both the optimal sensitivity and the optimal evolution time for the squeezing parameter and the QFI. In Fig.~\ref{fig:supp1}(b), we similarly show that DTWA simulations of squeezing and QFI under the gap-protected countertwisting Hamiltonian $\Hctg$ for $N=16$ spins with power-law interactions ($\alpha=6$) on a one-dimensional lattice agree well with exact diagonalization across a range of gap protection strengths $\chi^{-1}$.

\section{Spin-wave theory of squeezing}
\label{sec:supp_spin_wave}

The spin-wave theory introduced in the main text provides a broadly applicable framework for analyzing the early-time squeezing dynamics of interacting spin systems.  In particular, a generic XYZ Hamiltonian can be decomposed into terms that squeeze via parametric resonance (countertwisting) and terms that shift this process off resonance, inducing a phase space precession. Here, we elaborate on the derivation of the linear spin-wave theory, the resulting conditions for parametric instability, and the calculation of correlation functions to quantify metrological gain. While the main text applies these ingredients to analyze the early-time dynamics of gap-protected countertwisting, we here additionally show how spin-wave theory facilitates a comparison with other paradigmatic squeezing Hamiltonians.

\subsection{Spin-wave Hamiltonian and dynamics}
We consider the generic XYZ model in a $\uvec{z}$-field:
\begin{equation}\label{eq:HxyzSupp}
    H_\mathrm{XYZ} = \sum_{i,j,\mu} J_\mu(\vec{r}_{ij}) s^\mu_i s^\mu_j + h \sum_i s^z_i,
\end{equation}
where $\mu$\;$\in$\;$\left\{x,y,z\right\}$. Writing the same generic model in momentum space, with couplings $\tJ_\k^\mu $\;$\equiv$\;$ \sum_\vec{r} e^{-i \vec{k}\cdot \vec{r}} J^\mu(\vec{r})$ and spin operators $S^{\mu}_{\k} = \frac{1}{\sqrt{N}}\sum_{j} e^{-i \vec{k}\cdot\vec{r}_j}\vec{s}^{\mu}_j$, and expressing the transverse spin components $S^{x,y}$ in terms of spin raising and lowering operators $S^\pm$, we have:
\begin{equation}\label{eq:Hspinwave}
H_\mathrm{XYZ} = \sum_{\k} \left[\frac{J^x_\k - J^y_\k}{4} \left(S^+_\k S^+_{-\k} + S^-_\k S^-_{-\k}\right) + \frac{J^x_\k+J^y_\k}{4} \left(S^+_\k S^-_{-\k} + S^-_\k S^+_{-\k}\right) + J^z_\k S^z_\k S^z_{-\k}\right] + h \sqrt{N} S^z_0.
\end{equation}

The early-time dynamics of a system initialized in a spin-polarized state along the $\uvec{z}$ axis can be analyzed by applying a Holstein-Primakoff (HP) transformation, which maps the spin operators to bosonic operators $a^\dagger_{\k} $, $a_{\k}$ that create and annihilate spin waves of momentum $\k$ on top of the initial vacuum:
\begin{subequations}
\begin{align}
    S^+_{\k} &\approx a^\dagger_{-\k} \\
    S^-_{\k} &\approx a_{\k} \\ 
    S^z_{\k} &\approx -\frac{\sqrt{N}}{2}\delta_{\k,\vec{0}} + \frac{\tilde{n}_{\k}}{\sqrt{N}},
\end{align}
\end{subequations}
where $\tilde{n}_{\k}=\sum_{\vec{q}} a^\dagger_{\vec{q}}a_{\vec{q}+\vec{k}}$. The result of this transformation, up to quadratic order in the bosonic operators and assuming inversion symmetry ($J^\alpha_\k = J^\alpha_{-\k}$), is of the form:
\begin{equation}\label{eq:HhpSupp}
H_\mathrm{XYZ} \approx \sum_k \left[\chi_\k \left(\frac{a^\dagger_{-\k} a^\dagger_{\k} + a_\k a_{-\k}}{2}\right) + \omega_\k a^\dagger_\k a_{\k} \right].
\end{equation}
Here, the first term excites opposite-momentum spin-wave pairs, leading to squeezing of momentum modes $\pm\k$ in the $xy$-plane, at a rate
\begin{equation}\label{eq:chik}
\chi_\k = \frac{J^x_\k - J^y_\k}{2}.
\end{equation}
The second term describes the energy cost of creating a spin wave of momentum $\k$, as parameterized by the dispersion relation
\begin{equation}\label{eq:omegak}
\mathcal{\omega}_\k = \frac{J^x_\k+J^y_\k}{2} + h - J^z_0.
\end{equation}
For the case of gap-protected countertwisting, with $J^{x,y,z}_\k = (1+\chi, 1-\chi, 1)f_\k$ and $h=0$, Eqs.~\ref{eq:chik}-\ref{eq:omegak} yield squeezing rates $\chi_k = \chi f_\k$ and mode frequencies $\omega_\k = f_\k -1$, where we have fixed $f_0$=1 so that the energy scales extensively.  More generally, Eq.~\ref{eq:HhpSupp} provides a useful framework for analyzing the squeezing dynamics of an arbitrary XYZ spin model, by breaking it up into terms that squeeze via parametric amplification (at a rate set by $\chi_{\k}$) and terms that induce a phase space precession (at frequency $\omega_{\k}$) that detunes the parametric resonance.

The time evolution under the spin-wave Hamiltonian in Eq.~\ref{eq:HhpSupp} can be described analytically, by diagonalizing it via a Bogoliubov transformation,
\begin{equation}
    H_\mathrm{XYZ} \approx \sum_{\k} \lambda_{\k}\beta^\dagger_{\k}\beta_{\k}.
\end{equation}
Here, the eigenenergies
\begin{equation}\label{eq:lambdak}
\lambda_\k = \sqrt{\omega_\k^2-\chi_\k^2}
\end{equation}
are associated with Bogoliubov modes $\beta_\k$ given by
\begin{equation}
a_{\k}=\frac{\sqrt{\frac{\omega_{\k}}{\lambda_{\k}}+1}\,\beta_{\k}-\sqrt{\frac{\omega_{\k}}{\lambda_{\k}}-1}\,\beta^\dagger_{-\k}}{\sqrt{2}}.
\end{equation}
The Heisenberg equations of motion then take the simple form $\dot\beta_{\k}=-i\lambda_{\k}\beta_{\k}$.  Transforming the solution $\beta_\k(t) = e^{-i\lambda_\k}\beta_\k(0)$ to the basis of the original spin-wave modes gives:
\begin{subequations}
\begin{align}
a_\k(t) & =u_\k(t)a_\k(0) + v_\k(t)a^\dagger_{-\k}(0) \text{ , with coefficients} \\
u_\k(t) &= \cos{(\lambda_{\k}t)}-i\frac{\omega_{\k}}{\lambda_{\k}}\sin{(\lambda^+_{\k}t)} \\
v_\k(t) &= -i\frac{\chi_{\k}}{\lambda_{\k}}\sin{(\lambda_{\k}t)}.
\end{align}
\end{subequations}

We use this to compute the early-time pairwise connected correlators along the antisqueezed/squeezed quadratures $S^{\min/\max}=S^{\pm\pi/4}=(S^x\pm S^y)/\sqrt{2}$ of the gap-protected countertwisting Hamiltonian:
\begin{subequations}
    \begin{align}
    C_{ij}^{\text{min}/\text{max}}(t)&= \left\langle \left( \frac{s_i^x(t) \pm s_i^y(t)}{\sqrt{2}} \right) \left( \frac{s_j^x(t) \pm s_j^y(t)}{\sqrt{2}} \right)\right\rangle - \left\langle\frac{s_i^x(t) \pm s_i^y(t)}{\sqrt{2}} \right\rangle \left\langle\frac{s_j^x(t) \pm s_j^y(t)}{\sqrt{2}} \right\rangle \\
    &= \frac{1}{2N}\sum_{\k} e^{i \k\cdot \vec{r}_{ij}} \left( \left\langle (S_{\k}^x(t) \pm S_{\k}^y(t))(S_{-\k}^x(t) \pm S_{-\k}^y(t))\right\rangle - \left\langle S_{\k}^x(t) \pm S_{\k}^y(t) \right\rangle \left\langle S_{-\k}^x(t) \pm S_{-\k}^y(t) \right\rangle \right) \\ 
    &= \frac{1}{4N} \sum_{\k} e^{i \k\cdot \vec{r}_{ij}} \left( \left\langle a_{\k}(t) a^{\dagger}_{\k}(t) \right\rangle  + \left\langle a^{\dagger}_{-\k}(t) a_{-\k}(t) \right\rangle  \pm \left\langle i a_{\k}(t) a_{-\k}(t)\right\rangle  \mp \left\langle i a^{\dagger}_{-\k}(t)a^{\dagger}_{\k}(t)\right\rangle \right),
\end{align}
\end{subequations}
with initial conditions $\langle a_{\k}(0)a^\dagger_{\k}(0)\rangle=1$ and $\langle a^\dagger_{\k}(0)a_{\k}(0)\rangle=\langle a^\dagger_{\k}(0)a^\dagger_{\k}(0)\rangle=\langle a_{\k}(0)a_{\k}(0)\rangle=0$ corresponding to the vacuum state. In Fig.~\ref{fig:fig2}(b) of the main text, we show $C_{ij}^{\text{max}}$ and discuss its relation to the quantum Fisher information.

\subsection{Comparison of squeezing Hamiltonians}

The Bogoliubov eigenenergies $\lambda_{\k}$ in Eq.~\ref{eq:lambdak} impose a stability condition $|\omega_{\k}/\chi_{\k}|\ge1 $ for each momentum mode $\k$, since real (imaginary) eigenenergies indicate stability (instability) to parametric amplification. The ideal conditions to achieve fast, collective squeezing involve an unstable $\k=\vec{0}$ mode that squeezes exponentially fast; and stable $\k \neq \vec{0}$ modes, the excitation of which is energetically suppressed. 

In Table~\ref{tab:spin_wave}, we apply this formalism to comparatively study squeezing in the $xy$-plane under various XYZ Hamiltonians (Eq.~\ref{eq:HxyzSupp}) that have been examined in prior works~\cite{muessel2015twist,perlin2020spin,comparin2022robust,comparin2022scalable,block2024scalable,bornet2023scalable,wu2025spin,hu2017vacuum,li2023improving}. We specify the form of the XYZ couplings by factorizing $\tilde{J}^{\mu}_{\k}=J^{\mu}f_{\k}$ into the spin anisotropy $J^\mu$ and momentum-dependence $f_\k$, with normalization $f_0=1$, and additionally allow for a field $h$ along $\uvec{z}$.  We consider the gap-protected countertwisting Hamiltonian $\Hctg$ with $(J^{x,y,z},h)=(1+\chi,1-\chi,1,0)$; the gap-protected one-axis twisting Hamiltonian $\Hxxz$, generated by an XXZ model~\cite{perlin2020spin,comparin2022robust,comparin2022scalable,block2024scalable,bornet2023scalable,wu2025spin} in a rotated basis with $(J^{x,y,z},h)$\;$=$\;$(1-\chi,1,1,0)$; and the transverse-field Ising model $\Htfi$ with $(J^{x,y,z},h)=(1,0,0,h)$.  Gap-protected countertwisting $\Hctg$ achieves resonant parametric amplification in the $\k = \vec{0}$ mode by setting $\omega_0/\chi_0=0$, and can suppress excitation of $\k\neq \vec{0}$ modes arbitrarily strongly by decreasing the squeezing rate $\chi$ to produce a large gap $\omega_\k/\chi_\k$. On the other hand, while the XXZ model can arbitrarily suppress excitation of $\k\neq \vec{0}$ modes, it also suppresses parametric excitation of the $\k = \vec{0}$ mode by a constant ratio $\omega_0/\chi_0 = 1$ of the detuning to the parametric coupling.  The transverse field Ising model has an unstable $\k=0$ mode at a specific value of the transverse field $h=-\frac{1}{2}$, which mimics early-time countertwisting dynamics~\cite{muessel2015twist,hu2017vacuum,li2023improving} with weak gap protection $\chi^{-1}=1$.  However, $\Htfi$ lacks the capacity for tuning the strength of gap protection to scale the metrological gain attained by local interactions, a key feature of the model $\Hctg$ that is the focus of our work.

\begin{table}[h!]
    \centering
    \includegraphics[width=\textwidth]{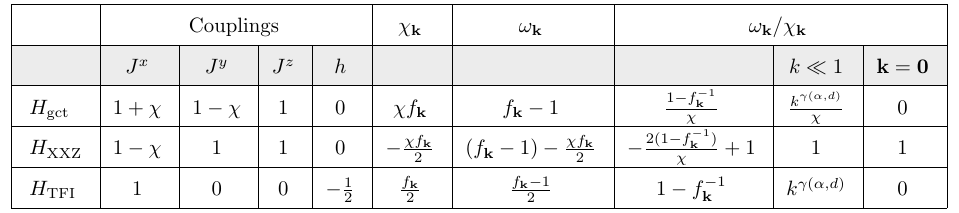}
    \caption{\textbf{Spin-wave analysis of squeezing dynamics: comparing Hamiltonians.} Parametric coupling $\chi_\k$ and energy cost $\omega_\k$ of exciting spin waves of momentum $\k$ under the gap-protected countertwisting Hamiltonian $\Hctg$, the gap-protected one-axis twisting model $\Hxxz$, and the Ising model $\Htfi$ with transverse field $h=-1/2$ tuned to mimic countertwisting dynamics. Also shown is the ratio $\omega_\k/\chi_\k$, together with its value for $\k=\vec{0}$ — which governs collective squeezing — and its behavior at small wavenumber $k \ll 1$ for power-law exponents $\alpha > d$ in $d$ dimensions. Only $\Hctg$ combines resonant parametric amplification ($\omega_\k/\chi_\k = 0$) with tunable gap protection $\omega_\k/\chi_\k \propto \chi^{-1}$ for suppressing excitation of modes with $k \neq 0$.}
    \label{tab:spin_wave}
\end{table}

\subsection{Gap protection in the spectrum}

In gap-protected countertwisting, the gap protection is achieved by energetically constraining dynamics to the permutation-symmetric manifold of collective spin $S=N/2$, i.e., the $\k=\vec{0}$ spin-wave mode. In previous sections, we used linear spin-wave analysis, which describes the early-time dynamical evolution from an initial spin-polarized state, to argue that this gap is set by the energy cost $\omega_{\k}=f_{\k}-1$ of exciting $\k\neq 0$ modes. In this section, we provide a detailed description of our calculation of the critical gap protection strength $\chi_c$ from spin-wave theory, and verify spin-wave predictions using exact diagonalization of small systems.

\begin{figure}[h]
\includegraphics[width=0.75\columnwidth]{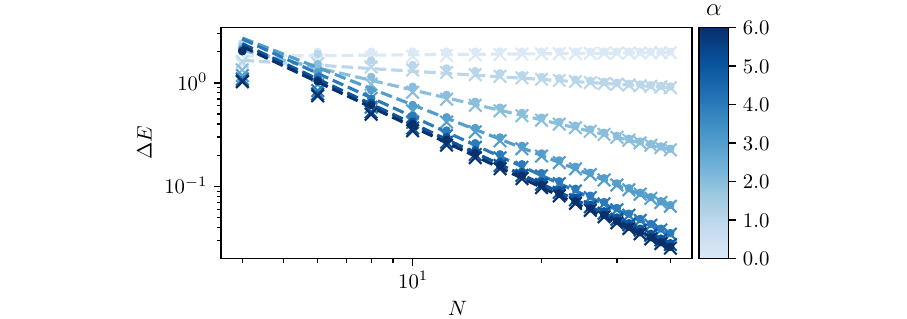}
\caption{\textbf{Spectral gap of Heisenberg Hamiltonian.} One-dimensional Heisenberg Hamiltonians $H_{\mathrm{Heis}}$ for $N\in[4,40]$ spins with various power-law exponents $\alpha$, we compute the energy gap $\Delta E_{\mathrm{gp}}$ between the spin-polarized state along the $\uvec{z}$ axis with quantum numbers $(J,m_z)=\left(\frac{N}{2},\frac{N}{2}\right)$, and the ground state of the sector with two excitations corresponding to quantum numbers $(J,m_z)=\left(\frac{N}{2}-2,\frac{N}{2}-2\right)$ (crosses). We overlay the energy cost $\Delta E_{\mathrm{sw}}\equiv 2\omega_{k_c}$ (circles) predicted by spin-wave theory for exciting a pair of lowest-lying excitations, where $k_c$ is the smallest nonzero wavenumber.}
\label{fig:supp2}
\end{figure}

For systems with periodic boundary conditions, due to translation invariance, the coupling matrix $f(\vec{r}_{ij})=J^{\mu}(\vec{r}_{ij})/J^{\mu}$ is diagonalized in momentum space, $f_{\k}$. The largest eigenvalue is the collective coupling $f_{\vec{0}}=1$, and the second largest eigenvalue is the coupling $f_{k_c}$ at the smallest nonzero wavenumber $k_c=2\pi/L$. To parametrically amplify the collective mode $\k=\vec{0}$ without exciting other modes, it is necessary to detune the nearest competing mode $k_c$, by enforcing $|f_{\vec{0}} - f_{k_c}| > \chi |f_{k_c}|$, so that it remains off-resonant. This sets the critical gap protection strength $\chi<\chi_c = 1-1/f_{k_c}$, which we evaluate numerically. 

We extend this analysis to systems with open boundary conditions, which break translation symmetry. In this case, we compute the two largest eigenvalues $\tilde{f}$ and $\tilde{f'}$ of the real-space coupling matrix, and analogously estimate the critical gap protection strength as $\chi < \chi_c = 1-\tilde{f}/\tilde{f'}$. Since these systems are nearly translationally invariant, the strongest mode $\tilde{f} \approx 1$ is approximately the permutation-symmetric mode, and the predictions agree well with DTWA results in Fig.~\ref{fig:fig2}(a).

Finally, we verify spin-wave predictions of the gap protection mechanism and scaling with system size, by evaluating the spectrum of the $N$-particle Heisenberg Hamiltonian
\begin{equation}
    H_{\mathrm{Heis}}=-\sum_{i\neq j}f(\vec{r}_{ij}) \vec{s}_i\cdot \vec{s}_j
\end{equation}
in one dimension. In particular, we compute the energy gap $\Delta E_{\mathrm{gp}}$ between the spin-polarized state along the $\uvec{z}$ axis with quantum numbers $(J,m_z)=\left(\frac{N}{2},\frac{N}{2}\right)$ and the ground state of the sector with quantum numbers $(J,m_z)=\left(\frac{N}{2}-2,\frac{N}{2}-2\right)$ corresponding to exciting a single pair of lowest-energy $\pm\k\neq 0$ modes. We plot these energy gaps as a function of system size in Fig.~\ref{fig:supp2} for power-law exponents $0 \leq \alpha \leq 6$, and verify that they agree with the spin-wave prediction of the cost $\Delta E_{\mathrm{sw}}\equiv2\omega_{k_c}=2(f_{k_c}-1)$ of exciting a pair with the smallest nonzero momentum $k_c = 2\pi/N$. By turning on perturbatively weak countertwisting at rate $\chi$, it is possible to achieve exponentially fast squeezing in the $\k=0$ mode while maintaining the energetic protection provided by $H_{\mathrm{Heis}}$ that penalizes excitations in the $\k\neq \vec{0}$ modes.

\section{System size scaling of optimal sensitivity and time}
\label{sec:supp_finitesize}

\subsection{Finite-size effects on sensitivity}

In the main text, we apply linear spin-wave theory to predict the critical squeezing rate $\chi_c$ to achieve Heisenberg scaling, as well as the metrological gain achievable when squeezing faster than $\chi_c$, examining the dependence on system size $N$, dimensionality $d$, and power-law exponent $\alpha$. In this section, we explore deviations to the predicted scalings due to finite-size effects via numerical simulations for a representative case $(d,\alpha)=(2,3)$.

\begin{figure}[h]
\includegraphics[width=0.75\columnwidth]{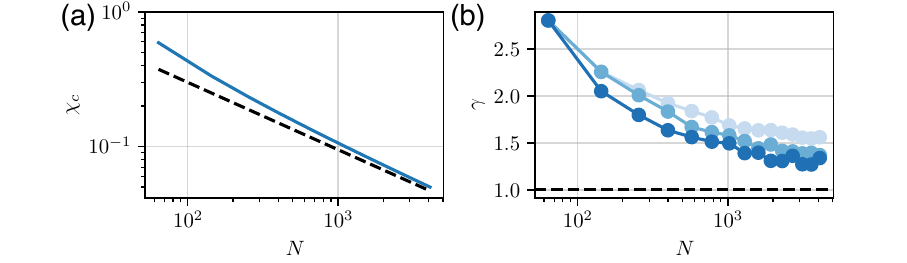}
\caption{\textbf{Finite-size effects.} \textbf{(a)} Critical gap protection strength $\chi_c = 1-1/f_{k_c}$ vs. $N$ in $(d,\alpha)=(2,3)$ (solid), where $k_c\propto N^{-1/d}$ is the smallest nonzero wavenumber in the $d$-dimensional lattice. At large $N$, this converges to the predicted scaling $
\chi_c \propto N^{-\gamma/d}$ with $\gamma(\alpha,d) = \min(2,\alpha-d) = 1$ (dashed). \textbf{(b)} $\gamma$ vs. $N$ extracted by fitting to the metrological gain associated with the QFI shown in Fig.~\ref{fig:fig2}(b) of the main text: $G=[N(\Delta\phiqfi)^{2}]^{-1} \propto \chi^{-d/\gamma}$, within the domains $\chi \in [\chi_c,1]$ (light), $\chi \in [\chi_c,4\chi_c]$ (medium), $\chi \in [\chi_c,2\chi_c]$ (dark). At large $N$, the fitted values approach the predicted $\gamma=1$.}
\label{fig:supp3}
\end{figure}

As a function of system size $N$, we predict the critical squeezing rate to scale as $\chi_c = 1-f_{k_c}^{-1} \propto N^{-\gamma(\alpha,d)/d}$ with $\gamma(\alpha,d)=\min(2,\alpha-d)$ for $\alpha > d$. This prediction is based on approximating the dispersion relation $\omega_\k = f_\k - 1$ by its behavior $\omega_\k \propto k^\gamma$ for $k\ll 1$ to estimate its value at the smallest nonzero wavenumber $k_c = 2\pi/N^{1/d}$.  In Fig.~\ref{fig:supp3}(a), we directly calculate $\chi_c = 1-f_{k_c}^{-1}$ by numerically evaluating $f_{k_c}$, and show that it converges to the predicted scaling for large system sizes.

On the other hand, for sub-critical gap protection $\chi^{-1} < \chi_c^{-1}$ --- i.e., for squeezing faster than the critical rate $\chi_c$ --- we predict the achievable metrological gain to scale as $G \propto (\chi_c/\chi)^{d/\gamma}$. In Fig.~\ref{fig:fig2}(b) of the main text, the scaling of the sensitivity $N(\Delta\phiqfi)^2 = 1/G$ with $\chi_c/\chi$ appears weaker than this prediction; however, the deviation is accounted for by finite-size effects. In particular, in two dimensions with power-law exponent $\alpha = 3$, the system exhibits linear dispersion $\gamma = 1$. For comparison, we fit the sensitivity obtained from the QFI in Fig.~\ref{fig:fig2}(b) as $N(\Delta\phiqfi)^2 = G^{-1} \propto (\chi_c/\chi)^{-2/\gamma}$ with $\gamma$ as a free parameter.  The fit value of $\gamma$, plotted in Fig.~\ref{fig:supp3}(b), decreases with $N$ and reaches $\gamma \approx 1.3$ for the largest simulated system size.

\subsection{Logarithmic scaling of optimal time}

\begin{figure}[h]
\includegraphics[width=0.75\columnwidth]{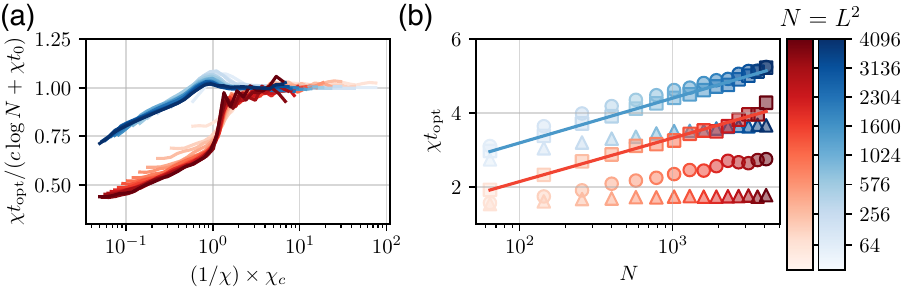}
\caption{\textbf{Optimal time scaling.} \textbf{(a-b)} Time $\chi t_{\mathrm{opt}}$ to achieve optimal squeezing (red) and QFI (blue) in $(d,\alpha)=(2,3)$, calculated with periodic boundary conditions for varying $N \le 4096$. The corresponding optimal sensitivities are plotted in Fig.~\ref{fig:fig2}(b-c) of the main text. \textbf{(a)} The optimal times collapse when rescaled by an offset logarithm $c\log{N}+\chi t_0$, as a function of gap protection $\chi^{-1}$ rescaled by the critical value $\chi_c^{-1}$ for each $N$. We extract $(c,t_0)=(0.51,-0.21)$ for squeezing and $(c,t_0)=(0.53,0.77)$ for QFI via the fitting procedure in (b). \textbf{(b)} Optimal time vs system size $N$, plotted at weak gap protection $\chi = 1$ (triangles), at criticality $\chi = \chi_c$ (circles), and at strong gap protection $\chi = 0.2 \chi_c$ (squares). We extract parameters $(c,\chi t_0)$ by fitting optimal times at strong gap protection ($\chi = 0.2 \chi_c$) to the function $\chi t_{\mathrm{opt}}=c\log{N}+\chi t_0$.
}
\label{fig:supp4}
\end{figure}

In this section, we verify the logarithmic dependence of optimal squeezing time $\chi t_\mathrm{opt}$, scaled by the squeezing rate $\chi$, on system size $N$. To this end, we provide supporting data for Fig.~\ref{fig:fig2}(b-c) of the main text, which presents DTWA simulations of the optimal sensitivities $N(\Delta\phi_\mathrm{sq/QFI})^2$ as functions of gap protection $\chi^{-1}$ and system size $N$. In Fig.~\ref{fig:supp4}, we show the corresponding optimal evolution times $\chi t_{\mathrm{opt}}$ for squeezing and QFI vs. $\chi^{-1}$ and $N$.

For sufficiently strong gap protection $\chi_c / \chi > 1$, the time required to reach optimal squeezing and QFI have a logarithmic dependence on system size, as evidenced by the collapse upon rescaling $\chi t_{\mathrm{opt}} / (c\log{N}+\chi t_0)$ in Fig.~\ref{fig:supp4}(a), as well as the linear dependence of $\chi t_{\mathrm{opt}}$ on $\log{N}$ for $\chi \ll \chi_c$ in Fig.~\ref{fig:supp4}(b).  This behavior matches our expectation that, for strong gap protection, the squeezing dynamics should approach that of the all-to-all model $\Hct$, where the logarithmic timescale arises from the exponential improvement in sensitivity $N(\Delta\phi)^2\propto e^{-2\chi t}$ towards the near-Heisenberg-limited saturation values $N(\Delta\phi_\mathrm{sq/QFI})^2=c_\mathrm{sq/QFI}/N$.

For sub-critical gap protection $\chi_c / \chi < 1$, where the attainable metrological gain is reduced, we correspondingly expect the optimal sensitivity to be reached at earlier time $\chi t_\mathrm{opt} \lesssim \log(N)$. The simulations in Fig.~\ref{fig:supp4} are consistent with this prediction.  The time to optimal QFI for $\chi_c / \chi < 1$ maintains a logarithmic dependence on system size, and the rescaled values $\chi t_{\mathrm{opt}} / (c\log{N}+\chi t_0)$ exhibit a linear scaling with $\log{(\chi_c/\chi)}$ up to $\chi_c / \chi \approx 1$ before saturating at criticality. On the other hand, time to optimal squeezing in the sub-critical regime is significantly shorter than for QFI  $\chi^{-1}<\chi_c^{-1}$, with a weaker than logarithmic dependence on system size, yet it increases sharply around $\chi \approx \chi_c$ before saturating at criticality.

\section{Echo Protocol for Gap-protected Countertwisting}
\label{sec:echo_protocol}

\begin{figure}[h]
\includegraphics[width=0.75\columnwidth]{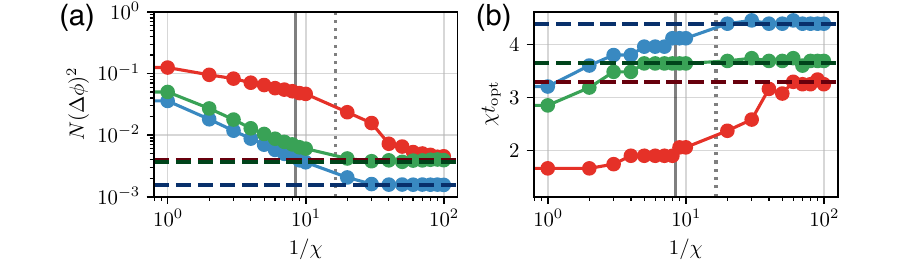}
\caption{\textbf{Echo protocol performance.} \textbf{(a)} Optimal sensitivity obtained from the echo protocol $N(\Delta\phiecho)^2$ (green), compared to squeezing parameter $N(\Delta\phisq)^2$ (red) and QFI $N(\Delta\phiqfi)^2$ (blue), with $N$\;$=$\;$1000$ spins in $(d,\alpha)=(2,3)$ with open boundary conditions, as a function of gap protection strength $\chi^{-1}$. Dashed curves show optimal sensitivity for all-to-all interactions ($\alpha = 0$). Grey lines indicate spin-wave prediction of critical value $\chi_c^{-1}$ for periodic (solid) and open (dotted) boundary conditions. \textbf{(b)} Optimal evolution times $\chi t_{\mathrm{opt}}$ corresponding to sensitivities in (a).}
\label{fig:supp5}
\end{figure}

The full metrological benefit of an entangled state is set by the quantum Cramér–Rao bound $N(\Delta\phi)^2 \geq N(\Delta\phiqfi)^2 = N/F$, where $F=4(\Delta \Smax)^2$ is the quantum Fisher information (QFI) with respect to the antisqueezed spin component. In approaching the Heisenberg limit, the squeezing parameter does not saturate the QFI, as it suffers from non-Gaussianity at late evolution times. As discussed in the main text, a concrete protocol for approaching the sensitivity set by the QFI is an echo sequence~\cite{davis2016approaching,anders2018phase,colombo2022time} consisting of evolution under $\Hctg$, rotation by a small angle $\phi \lesssim 1/N$ about the $\Smax$ axis, and subsequent time-reversed evolution under $-\Hctg$.  The reverse evolution amplifies the phase imprint into a large signal $\avg{\Smin}$ while recovering the noise of the initial coherent spin state, producing a sensitivity $N(\Delta\phiecho)^2 = \left[\partial_\phi \langle \Smin \rangle / S\right]_{\phi=0}^{-2}$. We plot the optimal sensitivities and the corresponding optimal times in Fig.~\ref{fig:supp5}, in comparison with the squeezing and QFI metrics. 

As shown in Fig.~\ref{fig:supp5}(a), while spin-wave theory accurately delineates the boundary of the region of large metrological gain by any measure, we observe the sharpest transition in sensitivity in the QFI and a slower transition in the squeezing. The echo protocol interpolates between the metrological gains associated with squeezing and QFI: the boundary $\chi_c$ aligns with QFI, whereas the saturation values agree with squeezing. As a result, the echo protocol is particularly helpful in approaching the full metrological gain set by QFI in the regime of weak gap protection $\chi^{-1}$. 

Fig.~\ref{fig:supp5}(b) elucidates the benefit of the echo protocol from the perspective of timescales. At small $\chi^{-1}$, where correlations remain short-ranged, the state quickly becomes non-Gaussian within a small collective Bloch sphere (of spin length $\abs{\avg{\vec{S}}} \ll N/2$).  Therefore, the squeezing parameter reaches a minimum well above the sensitivity set by the QFI at an early time $\chi t_\mathrm{opt}$.  The echo protocol provides increasing metrological gain out to longer time, thus yielding a clear advantage over squeezing. On the other hand, stronger gap protection $\chi^{-1}$ allows for generating longer-ranged correlations that set a larger collective Bloch sphere, within which non-Gaussianity appears at later times, allowing squeezing to approach the optimal value $N(\Delta\phisq)^2\approx 4/N=4\times10^{-3}$ of the all-to-all model.  Thus, for strong gap protection, the squeezing comes closer to the performance of the echo protocol, and all three measures of sensitivity approach the Heisenberg limit to within constant factors [Fig.~\ref{fig:supp5}(a)].

\section{Optimizing gap protection at fixed total interaction time}
\label{sec:supp_optimization}

Fig.~\ref{fig:fig4} of the main text showed the optimal sensitivity, together with the corresponding optimal squeezing rate $\chi$, as a function of total evolution time $\Jtot t$. Here, we elaborate on how this optimum is determined, and on the analytical model with which we compare the numerical simulations.

\begin{figure}[h]
\includegraphics[width=0.75\columnwidth]{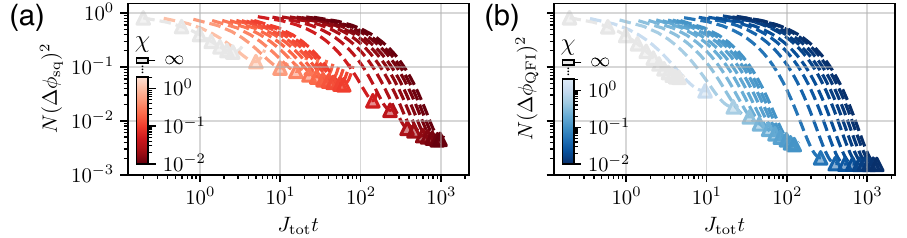}
\caption{\textbf{Optimization of squeezing rate.} \textbf{(a)} Squeezing $N(\Delta\phisq)^2$ (red) and \textbf{(b)} QFI $N(\Delta\phiqfi)^2$ (blue) versus total interaction time $\Jtot t$ for squeezing rates $0.01 \leq \chi \leq 2$ (dark to light) and $\chi=\infty$ (grey), simulated for $N=1000$ spins in $(d,\alpha)=(2,3)$ using DTWA. Markers show the sensitivity optimized over $\chi$ for each total interaction time $\Jtot t$, which traces the $\chi=\infty$ curve up to its optimum, and then follows the optimal metrological gain associated with each finite $\chi$.}
\label{fig:supp6}
\end{figure}

Since maximizing the metrological gain requires slowing down the squeezing rate $\chi$, in practice the optimal rate may be set by a constraint on the total evolution time.  Assuming that the countertwisting Hamiltonian is generated by Floquet engineering~\cite{liu2011spin,miller2024twoaxis}, the sum $J_\mathrm{tot} = \sum_\mu \abs{J_\mu} = \abs{1+\chi} + \abs{1-\chi} + 1$ of coupling strengths over Cartesian components $\mu$ determines the total evolution time $\Jtot t$.  We plot the dependence of squeezing and QFI on $J_\mathrm{tot} t$ in Fig.~\ref{fig:supp6}(a-b) for different squeezing rates $\chi$, focusing on $N=10^3$ spins in two dimensions with $\alpha=3$, simulated with DTWA.  The minima (triangles) of the curves indicate the optimum sensitivity given the total evolution time, which we summarize in Fig.~\ref{fig:fig4}(a) of the main text for different combinations $(d, \alpha)$ of dimensionality and power-law exponent. The corresponding optimal rates $\chi_\mathrm{opt}$ are shown in Fig.~\ref{fig:fig4}(b).

In Fig.~\ref{fig:fig4}(a) of the main text, we overlay the DTWA simulations with predictions from a simplified analytical model, as discussed in the main text and expanded further here. Our model predicts the dependence of optimal squeezing rate $\chi$ and the resulting optimal metrological gain on total interaction time $\Jtot t$ in three regimes: times $\Jtot t \lesssim 1$ too short to benefit from gap protection; intermediate times; and long times $\Jtot t \gtrsim 3 t_\mathrm{H}$ enabling sufficiently slow squeezing to approach the near-Heisenberg-limited performance of all-to-all countertwisting.

At small $\Jtot t$, the dominant limitation is the short evolution time. Therefore, the fastest possible squeezing $\chi \to \infty$ --- i.e., the limit of no gap protection --- is optimal.  The resulting sensitivity $N(\Delta\phi)^2 \approx e^{-2\chi t}$ at total evolution time $\Jtot t =2\chi t$ is shown by the solid grey curves in In Fig.~\ref{fig:fig4}(a). 

At intermediate $\Jtot t$, the short interaction range also starts to limit the maximum achievable gain. In this regime, the squeezing rate $\chi$ should be reduced to extend correlations and mimic all-to-all behavior over a larger volume $G\sim L_c^d$, at the expense of slowing down squeezing. The interplay between these two mechanisms determines the optimal value of $\chi$. In particular, at a given $\chi$, the early-time sensitivity improves exponentially fast with $\chi t$, before saturating a constant factor $c_{\mathrm{sq/QFI}}$ away from an effective Heisenberg limit associated with the volume of correlations $L_c^d\sim\chi^{-d/\gamma}$, which can be expressed simply as:
\begin{align}
   N(\Delta \phi_{\mathrm{sq/QFI}})^2 \approx \max{\left(e^{-2\chi t}, \frac{c_{\mathrm{sq/QFI}}}{L_c^d}\right)}.
\end{align}
We optimize the squeezing rate $\chi = 1-1/f_{2\pi/L_c}$ at fixed $\Jtot t$ by setting these two limits on the sensitivity equal to each other. The resulting predictions for the sensitivities $N(\Delta \phi_{\mathrm{sq/QFI}})^2$ are shown by the red/blue solid curves in Fig.~\ref{fig:fig4}(a).

Ultimately, decreasing the squeezing rate below the critical value $\chi_c$ associated with the total number of particles $N$ provides no benefit, as the sensitivities saturate to the ideal values $N(\Delta\phi_{\mathrm{sq}/\mathrm{QFI}})^2 $\;$=$\;$ c_{\mathrm{sq}/\mathrm{QFI}}/N$ achieved with the all-to-all countertwisting model $\Hct$, indicated by black solid lines in Fig.~\ref{fig:fig4}(a). We also indicate with black dashed lines the sensitivities achievable with the all-to-all twisting model $\Htwist$, which similarly sets the saturation values for optimal sensitivities obtained from the XXZ model.

\section{Robustness to fluctuating collective interactions}
\label{sec:supp_robustness}

In the main text, we examined the robustness of gap-protected countertwisting to fluctuating mean-field interaction strength, comparing with the case of XXZ models that realize gap-protected one-axis twisting.  A key distinction between countertwisting and one-axis twisting, present even in the collective spin models $\Hct$ and $\Htwist$, is whether the squeezed quadrature is fixed or rotates as a function of evolution time.  In particular, simulations in Fig.~\ref{fig:fig5}(a) showed that the fixed orientation of the squeezed quadrature under countertwisting $\Hct$ imparts robustness to fluctuating collective interaction strength, whereas the rotating squeezed quadrature under one-axis twisting $\Htwist$ leads the squeezing to saturate to a constant value at large system size $N$.  Here, we present the simple analytical model with which we compared the numerical simulations.

For reference, we first examine how a fluctuating collective interaction strength $\chi$ limits the squeezing attainable by applying one-axis twisting $\Htwist = \chi S_x^2/N$ to a coherent state initialized along $\uvec{z}$.  As a function of the strength $\chi t$ of squeezing, the orientation $\alpha$ of the squeezed quadrature $\Smin = \cos\alpha S^x + \sin\alpha S^y \equiv S^\alpha$ rotates, as illustrated in Fig.~\ref{fig:fig5}(a) of the main text. For intermediate times $1\ll \chi t \ll N^{1/3}$, where the metrological gain is large but the state remains Gaussian, the squeezed quadrature is at an angle $\alpha \approx 1/(\chi t)$ and has normalized spin noise $\zetamin \equiv 2\Delta \Smin/\sqrt{N} \approx 1/(\chi t)$.  A fractional change $\epsilon \ll 1$ in the mean-field interaction strength $\tilde{\chi} = (1+\epsilon)\chi$ perturbs the angle of the squeezed quadrature to
\begin{equation}
\tilde{\alpha} \sim \frac{1}{\chi t (1+ \epsilon)}\approx\alpha (1 - \epsilon).
\end{equation}
This rotation adds noise to the original squeezed quadrature by mixing in a small amount of the antisqueezing $\zetamax = 2\Delta \Smax/\sqrt{N} \approx \chi t$.  The resulting normalized spin noise
\begin{equation}
\widetilde{\zetamin^2} \approx \zetamin^2 + \epsilon^2 \alpha^2 \zetamax^2 \approx \frac{1}{(\chi t)^2}+\epsilon^2
\end{equation}
saturates to the constant value $\epsilon^2$ for large twisting strength $\chi t$.  We expect a similar saturation in sensitivity $N(\Delta\phi_\mathrm{sq})^2 = \widetilde{\zetamin^2}\left(S/\abs{\avg{S}}\right)^2$, since the spin length $\abs{\avg{S}}$ is unaffected by the rotation.  The simulation in Fig.~\ref{fig:fig5}(a) of one-axis twisting with rms fractional fluctuation $\Delta\chi/\chi = \epsilon$ in mean-field interaction strength is consistent with the sensitivity saturating to a value $N(\Delta\phi_\mathrm{sq})^2 \approx \epsilon^2$ at large $N$.

The countertwisting Hamiltonian $\Hct = \chi\left(S_+^2 + S_-^2\right)/(2N)$ avoids this saturation by squeezing along a fixed quadrature $\Smin = S^{\pi/4} = (S^x + S^y)/\sqrt{2}$.  This quadrature shrinks exponentially as a function of $\chi t$, with normalized spin noise $\zetamin \approx e^{-\chi t}$ for times $\chi t < \log N$ where the state remains Gaussian.  A small fractional change $\pm \epsilon$ in the collective interaction strength $\tilde{\chi}\approx\chi(1\pm\epsilon)$ affects the rate of squeezing, thereby modifying the normalized spin noise as
\begin{equation}
\widetilde{\zetamin^2}\approx e^{-2\chi(1\pm \epsilon)t}\approx \left(\zetamin^2\right)^{1\pm \epsilon}.
\end{equation}
Thus, the worst-case effect of a fractional change of magnitude $\epsilon\ll 1$ is to degrade the scaling of the normalized spin noise.  Further, for symmetric fluctuations $\pm \epsilon$, the increased noise $\widetilde{\zetamin^2}\approx \left(\zetamin^2\right)^{1- \epsilon}$ at decreased squeezing rate $\tilde{\chi}=\chi(1-\epsilon)$ dominates the scaling due to the nonlinear dependence of $\zetamin^2$ on $\tilde{\chi}$. 
For example, if the ideal squeezing at time $t$ is a fixed factor from the Heisenberg limit, with $\zetamin^2\propto 1/N$, then the scaling with $N$ will be degraded to $\widetilde{\zetamin}^2 \propto 1/N^{1-\epsilon}$.  The optimal sensitivity $N(\Delta\phi_\mathrm{sq})^2$ plotted in Fig.~\ref{fig:fig5}(a) is approximately consistent with this prediction, showing that the squeezing remains scalable even for fluctuating interaction strength.  In summary, because the countertwisting Hamiltonian preserves a fixed squeezed quadrature, it is markedly more robust to noise in the interaction strength than the one-axis twisting Hamiltonian in the limit of large metrological gain.

\section{Floquet engineering}
\label{sec:supp_floquet}

A demonstrated approach to realizing the gap-protected countertwisting Hamiltonian in systems with native XXZ interactions is Floquet engineering~\cite{miller2024twoaxis,gao2025signal}. Notably, the gap protection term can ensure that the couplings $J^{x,y,z}$ to be engineered are all of the same sign, in contrast to canonical two-axis countertwisting with $J^x = - J^y$. Thus, gap protection circumvents the need to change the sign of interactions during the Floquet pulse sequence, as first pointed out in a seminal proposal for transforming collective one-axis twisting to countertwisting~\cite{liu2011spin}. In this section, we derive the range of squeezing rates $\chi$ attainable for a given anisotropy of the native XXZ Hamiltonian, including the special cases of Ising and spin-exchange interactions, and numerically simulate the dependence of metrological gain on Trotter step size. 

Our objective is to convert native couplings $K^{x,y,z} = (K^{\perp}, K^{\perp}, \Delta K^{\perp})$ of the system Hamiltonian 
\begin{equation}
    H_{\mathrm{sys}} = \sum_{i,j,\mu} f(\vec{r}_{ij}) K^\mu s_i^\mu s_j^\mu
\end{equation}
into an average Hamiltonian $ \Hctg= \sum_{i,j,\mu} f(\vec{r}_{ij}) J^\mu s_i^\mu s_j^\mu$ with effective couplings $J^{x,y,z} = (1 + \chi, 1-\chi, 1)$. (We normalize $f_{\vec{0}}=\sum_{\vec{r}}f(\vec{r})=1$, consistent with the main text.) To engineer the average Hamiltonian, we use the framework for Floquet engineering and pulse sequence design presented in Ref.~\cite{choi2020robust}. The essence of the approach is to employ a rapid, time-periodic sequence of global $\pi/2$ pulses around the $\uvec{x},\uvec{y}$ axes, which toggle between different frames $\pm\mu=\pm (x,y,z)$ tracking the operator $\pm S^{\mu}$ that $+S^z$ transforms into.

\begin{figure}[ht]
\includegraphics[width=\columnwidth]{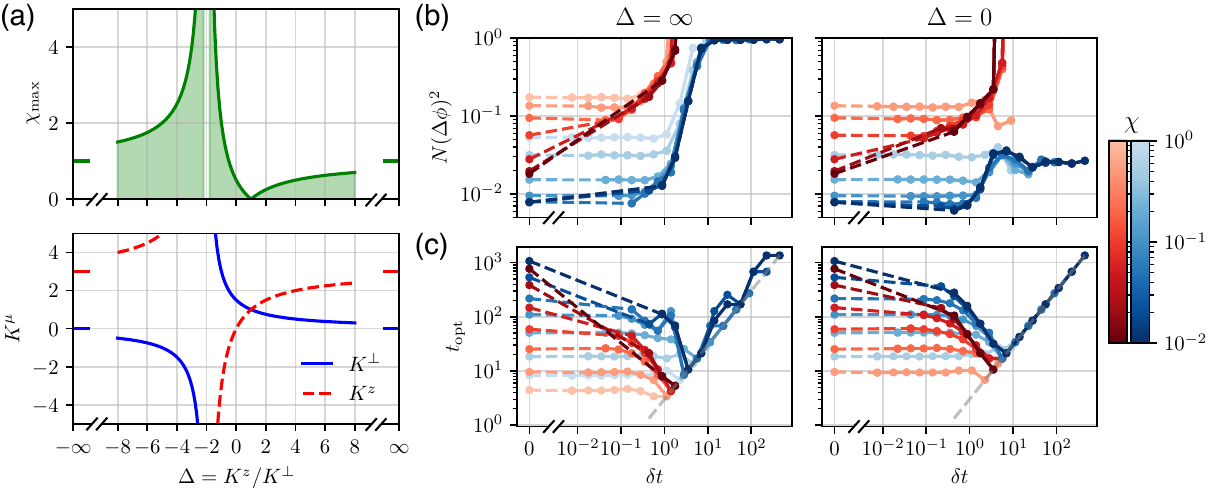}
\caption{\textbf{Trotterized implementation.} \textbf{(a)} Trotterizing a native XXZ Hamiltonian with anisotropy $\Delta = K^z/K^{\perp}$ enables realizing $\Hctg$ within the accessible range of squeezing rates $0 \leq \chi \leq \chi_{\max}$ shaded in green, and requires native couplings $K^{\perp,z}$ (blue, red). \textbf{(b)} Optimal sensitivity obtained from squeezing $N(\Delta\phisq)^2$ (red) and QFI $N(\Delta\phiqfi)^2$ (blue) under the trotterized implementation of $\Hctg$, as a function of Trotter step size $\delta t$, for $\Delta=\infty$, $\chi=1,0.5,0.2,0.1,0.05,0.02,0.01$ (light to dark), and for $\Delta=0$, $\chi=0.5,0.2,0.1,0.05,0.02,0.01$ (light to dark), for $N=200$ spins in $(d,\alpha)=(2,3)$. \textbf{(c)} Optimal evolution times $t_{\mathrm{opt}}$ corresponding to sensitivities in (b). The dashed grey curve marks a single Trotter step $t_{\mathrm{opt}} = 3 \delta t$. In the limit of fine trotterization $\delta t \rightarrow 0$, the Floquet evolution for time $t$ is equivalent to continuous evolution under $\Hctg$ for time $t'=t/3$, which we use to plot the sensitivity and optimal time $t_{\mathrm{opt}} = 3 t'_{\mathrm{opt}}$ at $\delta t = 0$ in (b-c).}
\label{fig:supp7}
\end{figure}

In this toggling frame picture, the fractional time $L_\mu$ that needs to be spent in the $+\mu$ or $-\mu$ frames depends on the target gap protection strength $\chi$, as well as the native spin-exchange coupling strength  $K^{\perp}$ and the native anisotropy $\Delta$:
\begin{subequations}
    \begin{align}
    L_x K^z + (1-L_x) K^{\perp} &= K^{\perp} (L_x \Delta + (1-L_x)) = (1+\chi) \\ 
    L_y K^z + (1-L_y) K^{\perp} &= K^{\perp} (L_y \Delta + (1-L_y)) = (1-\chi) \\ 
    L_z K^z + (1-L_z) K^{\perp} &= K^{\perp} (L_z \Delta + (1-L_z)) = 1.
\end{align}
\end{subequations}
Enforcing the constraint $\sum_\mu L_\mu = 1$, we solve for the required native spin-exchange coupling strength $K^{\perp}$ [Fig.~\ref{fig:supp7}(b)] and the fractional times $L_{\mu}$:
\begin{subequations}
\begin{align}
    K^{\perp}&=\frac{3}{\Delta+2}\\
    L_x &= \frac{1}{3} + \frac{\chi}{3} + \frac{\chi}{\Delta-1} \\
    L_y &= \frac{1}{3} - \frac{\chi}{3} - \frac{\chi}{\Delta-1} \\
    L_z &= \frac{1}{3}.
\end{align}
\end{subequations}
Furthermore, as global $\pi/2$ pulses do not change the sign of interactions, we use $L_\mu \geq 0$ to obtain upper bounds for the squeezing rate $\chi\geq 0$, defined to be non-negative:
\begin{equation}
    \chi \leq \chi_{\max}=\frac{1-\Delta}{2+\Delta} \cdot \begin{cases}
        1, & -2<\Delta \leq 1,\\
        -1, & \Delta<-2 \text{ or }\Delta>1.
    \end{cases}
\end{equation}

The range of accessible squeezing rates $\chi$ as a function of anisotropy $\Delta$ is depicted by the green shaded area in Fig.~\ref{fig:supp7}(a).  We observe that the regime $\chi \ll 1$ of interest for strong gap protection is accessible for almost any value of the anisotropy $\Delta$.  The only exceptions are the isotropic Heisenberg point $\Delta = 1$, for which the initial spin-polarized state is an eigenstate that undergoes no dynamics, and the limit $\Delta \rightarrow -2$.  For $\Delta = -2$, the isotropic component of the Hamiltonian, i.e. the Heisenberg term, averages to zero, or equivalently the required $K^{\perp}$ diverges. However, it is still possible to engineer the $\chi\to\infty$ Hamiltonian (no gap protection) by setting $(L_x,L_y,L_z)=(0,2/3,1/3)$, with native spin-exchange coupling strength $K^{\perp}=\chi$.

We provide explicit examples of the Floquet pulse sequences for two different cases, namely, native Ising or native spin-exchange (XY) couplings.  We construct pulse sequences with average Trotter step size $\delta t$ and Floquet period $\tau=\sum_\mu\tau_\mu=3\delta t$ that spend time $\tau_{\mu}=\delta t L_{\mu}$ sequentially in each frame $\mu=x\to z \to y$. If the native interactions are Ising interactions $\Delta \rightarrow \infty$, then $\Hctg$ can be engineered for $\chi \leq 1$, as seen in Fig.~\ref{fig:supp7}(a), through a periodic pulse sequence that spends time $(\tau_x,\tau_z,\tau_y)=\delta t \left(\frac{1+\chi}{3},\frac{1}{3},\frac{1-\chi}{3}\right)$ in each frame. Similarly, if the native interactions are symmetric spin-exchange interactions $\Delta = 0$, then $\Hctg$ can be engineered for $\chi \leq0.5$ through a periodic pulse sequence with $(\tau_x,\tau_z,\tau_y)=\delta t \left(\frac{1-2\chi}{3},\frac{1}{3},\frac{1+2\chi}{3}\right)$.

In Fig.~\ref{fig:supp7}(b-c), we examine how the optimal sensitivity and evolution time depend on the average Trotter step size $\delta t$, for a representative case of $N=200$ spins in $(d,\alpha) = (2,3)$. With native Ising interactions ($\Delta=\infty$), achieving metrological gain ($N(\Delta\phi)^2<1$) requires $\delta t \lesssim 1$. Below this threshold, the QFI (blue) saturates to its ideal value under continuous evolution, and can be enhanced by employing stronger gap protection $\chi^{-1}$ and longer evolution times. However, attaining enhanced squeezing (red) from stronger gap protection demands a correspondingly finer trotterization $\delta t$, and the Trotter step size for which the squeezing saturates decreases with increasing gap protection. 

While native XY interactions ($\Delta=0$) exhibit similar qualitative features, the QFI exhibits metrological gain even for $\delta t \gtrsim 1$. This increased resilience reflects the fact that even a single Trotter step of the XY Hamiltonian in the $\mu=x$ frame implements gap-protected one-axis twisting with substantial metrological gain~\cite{perlin2020spin}. However, squeezing does not benefit from a single Trotter step of size $\delta t \gtrsim 1$, as the squeezed state initially generated in the $\mu=x$ frame is distorted into a non-Gaussian state by the subsequent evolution in the $\mu=z$ frame.

% \bibliography{ctgap}